\documentstyle[preprint,aps]{revtex}

\newcommand{\be}{\begin{equation}}
\newcommand{\ee}{\end{equation}}
\newcommand{\ba}{\begin{eqnarray}}
\newcommand{\ea}{\end{eqnarray}}

\begin{document}

\title{Reconstructing Generalized Exponential Laws by Self-Similar 
Exponential Approximants}
\author{S.Gluzman$^1$, D. Sornette$^{1,2,3}$ and V.I. Yukalov$^{4,5}$}

\address{
$^1$ Institute of Geophysics and Planetary Physics \\
University of California, Los Angeles, California 90095 \\
$^2$ Department of Earth and Space Science\\
University of California, Los Angeles, California 90095 \\
$^3$ Laboratoire de Physique de la Mati\`ere Condens\'ee \\
CNRS UMR6622  and Universit\'e des Sciences \\
Parc Valrose, 06108 Nice Cedex 2, France \\
$^4$ Research Center for Optics and Photonics \\
Instituto de Fisica de S\~ao Carlos, Universidade de S\~ao Paulo \\
Caixa Postal 369, S\~ao Carlos, S\~ao Paulo 13560-970, Brazil \\
$^5$ Bogolubov Laboratory of Theoretical Physics \\
Joint Institute for Nuclear Research, Dubna 141980, Russia}

\maketitle

\begin{abstract}

We apply the technique of self-similar exponential approximants
based on successive truncations of continued exponentials to
reconstruct functional laws of the
quasi-exponential class from the knowledge of only a few terms
of their power series. Comparison with the standard
Pad\'e approximants shows that, in general, the self-similar exponential
approximants provide significantly better reconstructions.

\end{abstract}

\vskip 2cm

\section{Introduction}

Exponential laws are ubiquitous in nature. The overwhelming majority 
of relaxation phenomena occur through exponential laws. The dominant 
role played by exponential laws results from a combination of mechanisms.
First, exponential relaxations result from viscous dissipation proportional
to the first-order time derivative of the dynamical quantity, which is
usually the dominant term breaking the time-reversal symmetry (see however
\cite{radi} for a case where the third-order derivative becomes important).
Second, exponential decay or growth reflects the first-order expansion in the
rate of change as a function of the observable. Third,
exponential laws are often associated with the Poisson process, which has
the unique property of being memoryless. This leads to a remarkable
mathematical property: the Poisson law is invariant with respect to any
conditioning on the past or future and can thus be seen as the unique 
fixed-point of arbitrary transformations of time distributions involving 
conditioning \cite{sorkno}.

These universal properties provide useful benchmarks against which 
deviations can be gauged. These departures usually betray specific mechanisms at 
work in real systems, that are otherwise hidden by the just-mentioned universal mechanisms.
This is why the study of such laws which are close to exponentials is so important.
It may happen, however, that we are not able to observe in some 
experiment or to study theoretically such a law, say a relaxation process, from its beginning
till its very end. In theoretical works, this repeatedly happens when 
considering complicated problems that do not allow for exact solutions. Then, invoking
some kind of perturbation theory, one may calculate a few successive
approximations. The standard situation is when one employs a short-time
expansion resulting in approximations presented as polynomials over time.
The main problem in such cases is how to reconstruct the overall process
from the knowledge of only its short-time behavior represented by a series
in powers of time. An analogous situation may also arise in experiments, when one cannot, 
because of some technical difficulties, continuously measure the whole temporal
process, but one is able to gain information only from a limited number
of measurements performed at discrete times.
In that case again, one often presents the results in the simplest
form of a polynomial describing the given set of discrete points.

It may also happen that the whole relaxation process is impossible to 
observe experimentally because the lifetime of a system is shorter than 
the relaxation time of the process being studied. For instance, 
intensive investigations of various physical processes are now being 
accomplished for trapped atoms (see reviews \cite{1}-\cite{3}). Since 
the life time of atoms in a trap is finite, not all processes can be 
observed in full, in particular phenomena associated with trapped 
spinor Bose condensates \cite{4,5}.

One more example of the same problem arises when the characteristic 
relaxation times are too long to allow for convenient experiments 
covering the full range of the relaxation. Let us mention the 
spin-lattice relaxation in polarized nuclear magnets \cite{6,7}, in 
which the relaxation time may attain several days at low temperatures. 
Another illustration is the magnetic relaxation in molecular magnets 
\cite{8}-\cite{10}, where the relaxation time ranges up to several
months below the blocking temperature. There are cases when the 
characteristic times become comparable to or larger than human time 
scales, as for instance in visco-elastic relaxation in the earth crust
after earthquakes \cite{Shen,Massonnet}. Here, seismologists may only 
observe the early part of the relaxation process which may last for 
decades to centuries. In general, if one extracts information only 
from the beginning of a process, the result can again be presented 
as a power series. And the question that arises is how to find the 
general relaxation law from the knowledge of only its short-term 
presentation.

Here, we suggest a technique to address this question of how
to derive the whole sought function based on a perturbative expansion 
for short-time dynamics. To this end, we exploit the remarkable property
of self-similar exponential approximants introduced in  Ref.~\cite{11}, 
which is able to reconstruct exactly the exponential function $\exp(-t)$
from the knowledge of only the two first terms of its Taylor expansion.
In other words, the technique of self-similar exponential approximants
represents an ideal filter for the exponential function. It is thus 
natural to expect that this technique can be successfully applied for 
the reconstruction of continuous functions deviating from a pure 
exponential while still keeping the same exponential asymptotics at 
infinity, from the sole knowledge of their Taylor expansion for short 
times. We note that the self-similar exponential approximants generalize
and give a systematic justification of the method of summation of
power series by continued exponentials proposed by Bender and Vinson
\cite{benderwin} as an alternative to Pad\'e techniques, which
themselves the result of truncated continued fraction representations
of power series.

In the next section, we briefly review the main properties of the 
technique of self-similar exponential approximants that will be
useful for our purpose. In section 3, we then discuss several examples 
illustrating its power to reconstruct quasi-exponential laws. In section
4, we apply the technique to an ordinary nonlinear differential equation
motivated by a solid-friction problem. In each case, we compare our 
results with those from the more standard Pad\'e approximants. In 
general, we find that the numerical errors of the reconstructions are
significantly smaller for self-similar exponential approximants than 
for the standard Pad\'e approximants.

\section{Self-Similar Exponential Approximants}

The complete mathematical foundation of the method can be found in
Refs.~\cite{11}, \cite{12}-\cite{16}. Here, we give a brief sketch 
which emphasizes the variant that is mostly appropriate for filtering 
exponential-type laws. Assume that we are interested in a function 
$\phi (t)$ of a real variable $t$. Let perturbation theory (or some 
fitting procedure) give for this function the perturbative approximations
$\phi _n(t)$, with $n=0,1,2,...$ enumerating the approximation order. 
Consider the case when the perturbative procedure results in a polynomial
\begin{equation}
\label{1}\phi _n(t)\simeq \sum_{k=0}^na_k\ t^k\;\qquad (t\rightarrow
0),\qquad n=0,1,\ldots
\end{equation}
Let us stress that the expansion (1) has, as a rule, no direct meaning 
if continued straightforwardly to the region of finite arbitrary $t$.
In theoretical physics, the problem of reconstructing the value of a 
function at some distant moment of time from the knowledge of its 
asymptotic expansion as $t\rightarrow 0,$ is called the renormalization 
or resummation problem. An analytical tool for the solution of this 
problem, called algebraic self-similar renormalization, has been 
recently developed \cite{11}, \cite{12}-\cite{16}. The polynomial 
representation (1) gives for the sought function the following $n$
polynomial approximations $p_i(t),\ i=0,2,...,n,$
\begin{equation}
\label{2}
p_0(t)=a_0,\quad p_1(t)=a_0+a_1t,\quad p_2(t)=p_1(t)+a_2t^2,\ldots\; ,
\quad p_n(t)=p_{n-1}(t)+a_nt^n.
\end{equation}
The algebraic self-similar renormalization starts by applying to the
approximations (\ref{2}) an algebraic transformation, thus defining a 
new sequence, $P_i(t,s)=t^sp_i(t),\ i=0,2,...,n,$ with $s\geq 0.$ This
transformation raises the powers of the series (1), (2), and allows us 
to take effectively into consideration a longer timespan of the system 
history. We shall use below the strongest form of such transformation 
occurring by taking formally the limit $s\rightarrow \infty $. It can 
be shown that this results in an exponential representation of the 
sought function. The next step of the self-similar approximation theory
\cite{17}-\cite{20} consists in considering the sequence of transformed
approximations, $P_i(t,s),$ as a dynamical system in discrete time 
$i=0,1,...,n$, that we call ``time-order'' to distinguish it from the 
time variable $t$. In order to define the system evolution in time-order,
it is convenient to introduce a new variable $\varphi$ and to define 
the so-called expansion function $t(\varphi,s)$ from the equation 
$P_0(t,s)=a_0t^s=\varphi$, which gives $t(\varphi,s)=\left( \varphi
/a_0\right) ^{1/s}$. This makes it possible to construct the
cascade of approximations $y_i(\varphi,s)\equiv P_i($ $t(\varphi,s),s)$.
Embedding this cascade into a continuous approximation flow, one can 
write the evolution equation in terms of the discrete time-order 
variable in the form of the functional self-similarity relation, 
$y_{i+p}(\varphi ,s)=y_i(y_p(\varphi ,s),s),$ which is also the 
necessary condition for the convergence of $P_i$. Already at this stage,
we can try to check the effectiveness of the algebraic transformation 
by calculating the so-called local multipliers,
\begin{equation}
\label{3}m_i(t,s)\equiv \left[ \frac{\partial y_i(\varphi .s)}{\partial
\varphi }\right] _{\varphi =P_0(t,s)}\ ,
\end{equation}
as $s\rightarrow \infty $. When all $\left| m_i(t,\infty )\right| <1,$ the
convergence of the sequence $P_i$ is guaranteed. To implement
the calculations concretely, one can
use the integral form of the self-similarity relation,
$$
\int_{P_{i-1}}^{P_i^{*}}\frac{d\varphi }{v_i(\varphi ,s)}=\tau ,
$$
where the cascade velocity is 
$v_i(\varphi,s)=y_i(\varphi,s)-y_{i-1}(\varphi,s)$
and $\tau$ is the minimal number of steps of the approximation
procedure needed to reach the fixed point $P_i^{*}(t,s)$ of the
approximation cascade. It is possible to find $P_i^{*}(t,s)$ explicitly and
to perform an inverse algebraic transform after which the limit $%
s\rightarrow \infty $ is to be taken. The first step of the self-similar
renormalization is completed. This procedure is then repeated as many times
as necessary to renormalize all polynomials which appear at the preceding
steps. Completing this program, we come to the following sequence of
self-similar exponential approximations
\begin{equation}
\label{4}p_j^{*}(t,\tau _{1,}\tau _{2,}...,\tau _j)= a_0\exp \left( \frac{%
a_1 }{a_0}\tau _1 t\exp \left( \frac{a_2}{a_1}\tau_2 t...\exp \left( \frac{%
a_j}{a_{j-1}}\tau _jt\right) \right) ...\right) ,\quad j=2,3...,n.
\end{equation}
The quantities $\tau_1,\;\tau_2,\ldots $ play the role of control
functions that are to be defined from optimization conditions. For the
purpose of reconstructing an exponential-type law, the optimal choice of
controls $\tau_i$ is obtained by expanding $p_j^{*}(t,\tau _1,\tau 
_2,...,\tau _j)$
in the vicinity of $t=0$ and demanding that this expansion coincides
with $\phi_j(t)$. In the theory of Pad\'e approximants, this condition
is often refered to as the ``accuracy-through-order relationship.''
This method of determination of the control coefficients $\tau_i$
makes the self-similar exponential approximants analogous to the Euler
superexponentials \cite{benderwin,21,22}. With this determination
of the control parameters, we come
finally to the self-similar approximants $\phi_j^{*}$ of the sought
function,
$$
\phi _j^{*}(t,\tau _{1,}\tau _2,\ldots ,\tau _j)=p_j^{*}(t,\tau _1,\tau
_2,\ldots ,\tau _j)\;.
$$
For example,
$$
\phi _2^{*}(t,\tau _1,\tau _2)=a_0\exp \left( \frac{a_1}{a_0}\tau _1 t\exp
\left( \frac{a_2}{a_1}\tau _2 t\right) \right) ,
$$
$$
\phi _3^{*}(t,\tau _1,\tau _2,\tau _3)=a_0\exp \left( \frac{a_1}{a_0}\tau _1
t\exp \left( \frac{a_2}{a_1}\tau _2 t\exp \left( \frac{a_3}{a_2}\tau _3
t\right) \right) \right) ,
$$
$$
\phi _4^{*}(t,\tau _1,\tau _2,\tau _3,\tau _4)=a_0\exp \left( \frac{a_1}{a_0}%
\tau _1 t\exp \left( \frac{a_2}{a_1}\tau _2 t\exp \left( \frac{a_3}{a_2}\tau
_3 t\exp \left( \frac{a_4}{a_3}\tau_4 t\right) \right) \right) \right) .
$$
In order to check whether the sequence\ of $\phi _j^{*}(t,\tau _1,\tau
_2,...,\tau _j)$\ converges, we study their mapping multipliers, $%
M_j^{*}(t,\tau _1,\tau _2,\ldots \tau _j)$ defined as
\begin{equation}
\label{5}M_j^{*}(t,\tau _1,\ldots ,\tau _j)\equiv \frac{\delta \phi
_j^{*}(t,\tau _1,\ldots ,\tau _j)}{\delta p_1(t)}=\frac 1{a_1}\;\frac
\partial {\partial t}\;\phi_j^{*}(t,\tau _1,\ldots ,\tau _j)\;.
\end{equation}
This definition of the multipliers allows us to compare the convergence of the
expansion and of the renormalized expressions, making clear what can be
expected a priori.

With the control parameters defined as prescribed above for each $j$ from the
the accuracy-through-order relationship,  we
can obtain $j$ self-similar exponential approximants for the sought function
(where all $\tau $ are now known functions of the parameters $a_i$):
$$
\phi_{j1}^{*}(t)=\phi _j^{*}(t,\tau_1,1,...,1), \quad \phi
_{j2}^{*}(t)=\phi_j^{*}(t,\tau_1,\tau _{2},1,...,1), \quad
\phi_{jj}^{*}(t)=\phi_{j}^{*}(t,\tau _1,\tau _{2},...,\tau_{j}),
$$
which differ according to the number of control parameters being employed. This
provides a matrix of self-similar approximants, indexed by the order $j$ and
by the number of control parameters. To this matrix of approximants
is associated the matrix of multipliers. E.g., for $j=4,$
we have
$$
\phi _{21}^{*}(t)=\phi_j^{*}(t,\tau_1,1),\qquad \phi
_{22}^{*}(t)=\phi_j^{*}(t,\tau _1,\tau _{2}),
$$
$$
\phi _{31}^{*}(t)=\phi_3^{*}(t,\tau _1,1,1), \quad \phi
_{32}^{*}(t)=\phi_3^{*}(t,\tau _1,\tau _{2},1),\quad \phi _{33}^{*}(t)=\phi
_3^{*}(t,\tau_1,\tau_2,\tau_3),
$$
$$
\phi _{41}^{*}(t)=\phi _4^{*}(t,\tau_1,1,1,1)\; , \qquad
\phi_{42}^{*}(t)=\phi_4^{*}(t,\tau _1,\tau_2,1,1) \; ,
$$
$$
\phi _{43}^{*}(t)=\phi_4^{*}(t,\tau _1,\tau _2,\tau _3,1)\; , \qquad 
\phi_{44}^{*}(t)=\phi_4^{*}(t,\tau_1,\tau _2,\tau _3,\tau _4).
$$
We propose to qualify the convergence towards the sought function 
by examining both the convergence of the different sequences of
multipliers $M^{*}$ and of the approximants $\phi^*$, choosing among 
them the pair which exhibits the best convergence rate. The resulting 
limiting point from the table of approximants should be taken for the 
value of the sought function. If there are more than one limiting point,
one can take their weighted average, following Ref.~\cite{16}.

\section{Illustration of Technique in Action}

We now present explicit examples demonstrating how the technique
works.

\subsection{Ideal Filter}

Consider the Taylor expansion of the exponential function, $\phi (t)=\exp
(-t),$ up to an arbitrary order in $t$,
\be
\phi_n(t)\simeq \sum_{k=0}^na_k\ t^k\;,
\label{mgnkr}
\ee
where
$$
a_0=1,\quad a_1=-1,\quad a_2=\frac 12,\quad a_3=-\;\frac 16,\quad a_4=\frac
1{24},\quad a_5=-\;\frac 1{120}.
$$
The second-order self-similar approximant is
$$
\phi _{22}^{*}(t)=a_0\exp \left( \frac{a_1}{a_0}\;t\;\tau _1\exp \left(
\frac{a_2}{a_1}\;t\;\tau _2\right) \right) ,\qquad \tau _1=1,\quad \tau
_2=1-\;\frac 12\;\frac{a_1^2}{a_2a_0}=0.
$$
The fact that $\tau _2=0$ leads to $\phi _{22}^{*}(t)=\exp (-t)$. It also
makes any arbitrary order approximant $\phi _{jj}^{*}(t)$ identical to $\exp
(-t)$. Note, that all other approximants, except $\phi _{21}^{*}$, $\phi
_{31}^{*}$, $\phi _{41}^{*}$ and $\phi _{51}^{*}$, become identical to $\exp
(-t)$ and all higher order control parameters are identical zeros, 
like $\tau _2.$

Consider now the conventional Pad\'e approximants, P$_3^2(t)$ and P$_4^1(t)$
(P$_1^4(t)$ and P$_2^3(t)$ have inferior quality) and compare them to the
exact function, see Fig. 1. Although P$_3^2(t)$ remains positive, it behaves
nonmonotically. On the other hand, P$_4^1(t)$ becomes negative for large $t$.
These are typical problems encountered while attempting to reconstruct
functions with exponential asymptotic behavior by means of Pad\'e
approximants. The relative percentage errors for those two Pad\'e approximants
are shown in Fig. 2. One should not be mislead by the seemingly superior
performance of P$_4^1(t)$, which is qualitatively wrong in predicting
negative values already for moderate times. Because of this, we will
present below only the relative percentage error for positively defined Pad\'e
approximants.

Let us study the impact of noise on the coefficients of the power law expansion
(\ref{mgnkr}). Since $\phi_{22}^{*}(t)$ recovers the exact solution 
in absence of
noise, we study its perturbed value denoted  $\phi_{22}^{*}(t,\eta ,\theta)$
brought by the existence of the noises
$\eta$ and $\theta$ on the two first coefficients of the expansion 
(\ref{mgnkr})
defined by the replacement of
$a_1$ by $a_1(1+\eta )$ and/or of $a_2$ by $a_2(1+\theta)$.
For sufficiently small amplitudes of the
fluctuations, one can expand $\phi_{22}^{*}(t,\eta ,\theta )$ as
follows (only the most simple cases are considered):
$$
\phi _{22}^{*}(t,\eta ,0)-\phi _{22}^{*}(t)\simeq b_1(t)\ \eta +b_2(t)\ 
\eta^2+...,
$$
and
$$
\phi _{22}^{*}(t,0,\theta )-\phi _{22}^{*}(t)\simeq c_1(t)\ 
\theta +c_2(t)\ \theta ^2+...,
$$
where all coefficients can be easily calculated analytically. We
observe that:

1. the absolute error in the lowest order tends to peak at intermediate
times, see Fig 3, while the relative error monotonously increases with 
time;

2. the absolute error caused by fluctuations in $a_2$ is smaller than 
the error caused by fluctuations of equivalent magnitude in $a_1$, see 
Fig 3;

3. higher order terms in the expansion tend to shift the peak in the
absolute error towards larger times, see Fig. 4;

4. Interference of the error peaks at different orders leads to broaden the
domain in time over which the errors are significant, see Fig. 4.

It is possible to perform an averaging over the random variables and 
evaluate the
corresponding averages as a formal series to develop a specific resummation
procedure that may provide a solution for the average function, even for
arbitrary strong noises. We do not persue this here and return to the study
of how self-similar exponential
approximants, which are ideal filters for the exponential function,
extract information on other functions which are perturbations
of exponentials, from a few starting terms of their
Taylor expansions.

\subsection{Weak perturbation}

Consider the function, $\phi (t)=\sin (t)^2\exp (-t^2),$ which is
nonmonotonic. Its expansion is
\be
\phi_5(t)\simeq t^2\sum_{k=0}^5a_k\ t^{2k}\;,
\label{mmlls}
\ee
where
$$
a_0=1,\quad a_1=-\;\frac 43\;,\quad a_2=\frac{79}{90}\;,\quad a_3=-\;\frac
8{21}\;,\quad a_4=\frac{13921}{113400}\;,\quad a_5=-\;\frac{29341}{935550}%
\;.
$$
The coefficients alternate in sign, like in the ideal exponential 
example. The values of
the control parameters,
$$
\tau _2=0.013,\ \quad \tau _3=-0.137,\quad \tau _4=-0.131,\quad \tau
_5=-0.152,
$$
are all negative, with $\tau _2$ being rather small, which guarantees a good
quality of the approximation already in the lowest nontrivial order.
Higher-order approximants provide corrections to the good starting
approximation $\phi _{22}^{*}(t).$

Both multipliers and approximants converge extremely well. It is difficult
to distinguish in Fig. 5 the approximant $\phi_{55}^{*}(t)$ from the exact
function. The relative error at $t=2$ is $0.045\%$ for this approximant, while
$\phi_{5}(t=2)$ defined in (\ref{mmlls})
has a very large error $-4.058\times 10^5\;\%$. Note that the nondiagonal
forms $\phi_{41}^{*}$ and $\phi _{51}^{*}$ approximate the sought function
rather badly compared to the diagonal approximants.
The logarithms of the relative percentage errors for the Pad\'e
approximant P$_3^2(t)$ and for the super-exponential
approximant $\phi_{55}^{*}(t)$, corresponding
to the same number of terms used in the construction of both 
approximants, are shown in Fig. 6.

\subsection{``Badly damaged'' expansion}

Let us consider a function with an expansion whose
coefficients do not alternate in sign, such as $\phi (t)=\ln (1+t)\exp
(-t^2)$, which has the expansion
\be
\phi _5(t)\simeq t\sum_{k=0}^5a_k\ t^k\;,
\label{mbgnjnjkdk}
\ee
with the coefficients
$$
a_0=1\;,\quad a_1=-\;\frac 12\;,\quad a_2=-\;\frac 23\;,\quad a_3=\frac
14\;,\quad a_4=\frac{11}{30}\;,\quad a_5=-\;\frac 16\;.
$$
This is again a nonmonotonic function. The control parameters,
$$
\tau _2=1.187,\ \quad \tau _3=1.69,\quad \tau _4=-0.296,\quad \tau _5=0.661,
$$
have different signs, with $\tau _2$ and $\tau _3$ being rather large, which
leads to a poor quality of approximation to the exact function for 
$\phi_{22}^{*}(t)$ and $\phi
_{33}^{*}(t)$. Rather than finding a good approximation to the exact solution
already at the first step of the construction, the successive 
self-similar exponential
approximants bracket the exact function within
a rather broad range, as seen from the change of the sign
between $\tau _3$ and $\tau_4.$  The analysis
of the multipliers shown in Fig. 7 shows that
$M_{44}^{*}(t)$ and $M_{55}^{*}(t)$ are the closest, which suggests a
better convergence and that $\phi _{55}^{*}(t)$ and $\phi _{44}^{*}(t)$
are the two best approximants. Indeed, they are much
closer to the exact function than the lower-order approximants
as shown in Fig. 8.
Also, since $|M_{55}^{*}(t)|<|M_{44}^{*}(t)|$, one can anticipate
correctly that $\phi_{55}^{*}(t)$ is located closer to the exact 
solution than $\phi_{44}^{*}(t)$.

Consider all Pad\'e approximants (P$_3^2(t)$, P$_4^1(t)$, P$_1^4(t)$, 
and P$_2^3(t)$)
which can be built from the fifth-order
polynomial (\ref{mbgnjnjkdk}). One can see
in Fig. 9 that their quality is by far inferior to that obtained with 
$\phi_{55}^{*}$.
Fig. 10 shows the logarithm of the relative percentage errors for the
self-similar exponential approximant $\phi_{55}^{*}(t)$  and for the
Pad\'e approximant P$_1^4(t)$, which is the only one remaining positive.

\subsection{Strong perturbation}

Consider the Taylor expansion $\phi _5(t)\simeq \sum_{k=0}^5a_k\ t^k$
of the exponential-type function, $\phi
(t)=\frac 1{1+t}\exp (-t),$ up to the fifth order in $t$,
with the coefficients
$$
a_0=1,\quad a_1=-2,\quad a_2=\frac 52,\quad a_3=-\;\frac 83,\quad a_4=\frac{%
65}{24},\quad a_5=-\;\frac{163}{60}.
$$
The coefficients alternate in sign but their amplitudes behave
nonmonotically. The approximants $\phi _{22}^{*}(t)$, $\phi _{33}^{*}(t)$, $%
\phi _{44}^{*}(t)$, $\phi _{55}^{*}(t)$, and multipliers $M_{22}^{*}(t)$, $%
M_{33}^{*}(t)$, $M_{44}^{*}(t)$, $M_{55}^{*}(t)$ for the sought function can
be readily written down. The values of the control parameters,
$$
\tau _2=0.2,\ \quad \tau _3=0.508,\quad \tau _4=0.377,\quad \tau _5=0.378,
$$
are all positive and smaller than one.

Fig.11 shows the dependence of the multipliers as a function of time.
One can observe that $M_{44}^{*}(t)$ and $M_{55}^{*}(t)$ are the closest
pair, suggesting a good convergence of the cascade of approximants. The 
corresponding approximants, $\phi_{44}^{*}(t)$ and $\phi _{55}^{*}(t)$ 
bracket the exact function from above
and below respectively, see Fig. 12. One can observe that the
approximant $\phi _{33}^{*}(t)$ has already a rather good quality of 
approximation of the exact function. Relative percentage errors for 
the Pad\'e approximant P$_3^2(t)$ and for the self-similar exponential 
approximant $\phi_{55}^{*}(t)$ are shown in Fig. 13. In a narrow region, 
P$_3^2(t)$ outperforms $\phi_{55}^{*}(t)$ but fails short for larger 
times. However, due to the fact that the control parameters
$\tau_2$, $\tau_3$, $\tau_4$ and $\tau_5$ remain all positive and
of similar magnitude, the quality of $\phi_{55}^{*}(t)$ becomes eventually
worse than that of the Pad\'e approximation at very large times.

To investigate this behavior some more, let us
construct the following function, $\phi (t)=\exp (-t\
\left( 1+t\right) ^{-1/2}),$ which decays as $\exp (-\sqrt{t})$, as time
goes to infinity. Its Taylor expansion, $\phi _5(t)\simeq \sum_{k=0}^5a_k\
t^k\;$ possesses the coefficients
$$
a_0=1,\quad a_1=-1,\quad a_2=1,\quad a_3=-\;\frac{25}{24},\quad a_4=\frac{53
}{48},\quad a_5=-\;\frac{2261}{1920}.
$$
The control parameters,
$$
\tau _2=0.5,\ \quad \tau _3=0.48,\quad \tau _4=0.393,\quad \tau _5=0.367,
$$
are all positive, smaller than one but decrease rather slowly. Fig. 14
shows the dependence of the multipliers as a function of time.
One can observe their poor convergence. As a consequence,
the self-similar exponential approximants $\phi_{44}^{*}(t)$ and 
$\phi_{55}^{*}(t)$
bracket the exact
function rather poorly, see Fig. 15, with an accuracy inferior to that 
of the Pad\'e approximant P$_4^1(t)$, with the exception of very large 
times, see Fig. 16.

These two examples illustrate the property that, for coefficients 
$a_i$'s rapidly growing in absolute values, the considered
self-similar exponential approximants, with controls described by the
accuracy-trough-order relationship, become unreliable. Such cases 
of coefficients growing as fast as a factorial of their order 
constitute an important class of behavior, since it appears in
expansions that are typical to many nonlinear field theories. 
Consider for instance the particularly illustrative
example of the Stieltjes function, $\phi (t)=\int_0^\infty \frac{\exp
(-u)}{1+tu}du$, which exemplifies such a behavior. Its coefficients of
the Euler series \cite{24}
$$
a_0=1,\quad a_1=-1!,\quad a_2=2!,\quad a_3=-\;3!,\quad a_4=4!,\quad a_5=-5!,
$$
diverging as a factorial, lead to the control parameters
$$
\tau _2=0.75,\ \quad \tau _3=0.713,\quad \tau _4=0.697,\quad \tau _5=0.685~.
$$
As we can anticipate from our previous observations, both the 
multipliers and approximants sequences are not convergent 
simultaneously, as seen in Figs. 17 and 18. In this case, the Pad\'e 
approximant P$_3^2(t)$ easily outperforms $\phi_{55}^{*}(t),$ see 
Fig. 19.

We stress that this failure of the Euler-type self-similar exponential 
approximants, as compared to the Pad\'e approximants, does not imply 
the failure of the algebraic self-similar renormalization as a whole. 
Rather, it demonstrates the limitations of the particular way of 
defining controls and also of the infinite-power condition
($s\rightarrow \infty$) used in the derivation of these approximants.
In applications, the usage of the exponential approximants is often 
analogous to mean-field-type approximations. Because of this, we may
call, for brevity, the limit $s\rightarrow\infty$ as the mean-field
condition. It is important to realize that this mean-field theory sends 
clear warnings about its own anticipated failure, reflected in the 
divergent behavior of the approximants and of their multipliers. In a 
subsequent paper, we show that lifting of the infinite-power restriction
improves dramatically the accuracy of the self-similar renormalization.

\section{Exponential approximants derived from a differential equation}

We end our exploration of illustrative examples by the analysis of an expansion
derived from the time evolution equation of the state variable usually
called $\theta$ in the friction literature of a block
subjected to constant shear over normal stress, given by the Ruina-Dieterich
solid friction law (see \cite{booksor} page 283 section 13.6.2 and
\cite{Dieterich,Clapiere}). Posing $x\equiv \theta / \theta_0$ where $\theta_0$
is a parameter of the constitutive friction law, the following equation
\begin{equation}
\label{6}\frac{dx}{dt}=\theta _0^{-1}-\alpha \ x^{1-m},\qquad x(0)=x_0>0~,
\end{equation}
describes the evolution of the state variable of the friction law. In the
following, we shall consider the case when
the condition $\theta_0^{-1}-\alpha x_0^{1-m}<0$ holds which ensures that
the solid friction state variable tends to decrease from an initial 
large value.
This law (\ref{6}) results from a velocity-dependent solid friction 
coefficient,
the block velocity $V$ being related to the state variable $x$ by the relation
$V \propto 1/x^m$. $\alpha$ is a parameter of the constitutive 
velocity-dependent
solid friction law. The exponent $m$ is also a parameter dependent upon the
physical nature of the solid contacts.
Equation (\ref{6}) with initial condition
$\theta_0^{-1}-\alpha x_0^{1-m}<0$ thus describes the acceleration of a block
in contact with a solid substrate pulled with a constant force which has been
suddenly applied at $t=0$. It has been applied to describe one of the possible
regimes of a mountainous slope which can become transiently unstable 
\cite{Clapiere}.

In dimensionless variables, $X=x/\overline{x},\quad T=t/\overline{t}$, where
\be
\overline{x}=\left( \alpha \theta _0\right) ^{\frac 1{m-1}},\qquad \overline{%
t}=\left| 1-m\right| ^{-1}\alpha ^{\frac 1{m-1}}\theta _0^{\frac m{m-1}}~,
\ee
Eq.~(\ref{6}) reads,
\be
\frac{dX}{dT}\ \left( \left| 1-m\right| \right) =1-\ X^{1-m},\qquad
X(0)=X_0= \frac{x_0}{\overline{x}},\qquad 1-X_0^{1-m}<0~.
\ee
For $m<1$, it is a nonlinear relaxation equation, with $\overline{x}$ being
an equilibrium value reached asymptotically after an exponential decay
characterized by the typical time $\overline{t}$. For $m<1$, in a
long-time limit, we obtain an asymptotic solution in the form,
\be
x(t)\simeq \overline{x}+A_1\exp \left( -t/\overline{t}\right) +...,\qquad \
t\rightarrow \infty~,
\ee
where the value of $A_1$ remains unknown. A naive approach consists then in
writing down a naive
approximate solution to Eq.~(\ref{6}), $x_N(t)$, based on such an 
asymptotic single
relaxation time expression,
\be
x_N(t)=\overline{x}+(x_0-\overline{x})\exp \left( -t/\overline{t}\right) ~.
\ee
In this ansatz, the amplitude $A_1$ is now determined.

The exact expansion $x_k(t)$ up to order $t^k$ for short times and
arbitrary $m$ can be obtained from Eq.~(\ref{6}). Here, we limit ourselves to
expansion up to the fifth order,
\begin{equation}
\label{7}x_5(t)\simeq \sum_{n=0}^5a_nt^n,\ t\rightarrow 0,
\end{equation}
where in dimensional units,
$$
a_0=x_0,\quad a_1=\theta _0^{-1}-\alpha \ x_0^{1-m},\quad a_2=\frac 12\alpha
\ (m-1)\ a_1x_0^{-m},
$$
$$
a_3=\frac 16\alpha (m-1)\left[ -m\ a_1^2\ x_0^{-m-1}+2a_2\ x_0^{-m}\right] ,
$$
$$
a_4=\frac 1{24}\alpha (m-1)\left[ \left( 1+m\right) \ m\ a_1^3\
x_0^{-m-2}-6m\ a_2\ a_1x_0^{-m-1}+6a_{3\ }x_0^{-m}\right] ,
$$
$$
a_5=\frac 1{5!}\alpha (m-1)\left[
-m(1+m)(2+m)a_1^4x_0^{-m-3}+12m(1+m)x_0^{-m-2}a_1^2a_2-\right.
$$
$$
\left. -24mx_0^{-m-1}a_3a_1-12mx_0^{-m-1}a_2^2+24x_0^{-m}a_4\right] .
$$

Using this expansion, we apply our technique to construct a self-similar
exponential approximant using this fifth-order expansion (\ref{7}).
For $m=0.85,$ $X_0=50$, Fig. 20 illustrates how well the ``exact'' numerical
solution is approximated by superexponential approximants in dimensionless
units. The values of the control parameters,
$$
\tau _2=-2.02,\ \quad \tau _3=1.504,\quad \tau _4=0.848,\quad \tau _5=0.649,
$$
have different signs, with $\tau_2$ and $\tau_3$ being rather large, which
leads to a limited quality of the approximations at different orders. We
stress that these values for the model parameters correspond to a 
highly nonlinear and
strongly out-of-equilibrium case, where the naive exponential expression $%
x_N(t),$ corresponding to a close-to-equilibrium linearized model is way-off
the mark both in qualitative and quantitative sense, see Fig. 20. In order
to guarantee that the solution saturates at the correct constant value at
infinite time, we add and subtract $\overline{x}$ from the initial series,
which produces a single modification in that only
$a_0$ is replaced by $a_0-\overline{x}$ in the generic
expressions for the approximants. We thus plot
\be
x(t)=\overline{x}+x_{ii}^{*}(t),\ i=2,3,...5~,
\ee
divided by $\overline{x}$ as a function of the dimensionless time $T$.
Note that such shifts are completely compatible with the framework of 
algebraic self-similar renormalization, see e.g. Ref.~\cite{15}. For 
the fifth-order approximant studied here, the relative error is largest 
around $T=12.57$ and amounts to $1\%.$

In order to compare the accuracy of the self-similar approximants with 
the Pad\'e approximants \cite{24}, we consider the Pad\'e approximant,
\be
P_3^3(t)=\frac{a_0+A_1t+A_2t^2+B_3\overline{x}t^3}{1+B_1t+B_2t^2+B_3t^3}\ ,
\ee
which goes to $\overline{x}$ at infinity, with all unknown coefficients
determined from the short time expansion (\ref{7}). Fig.~21 shows that $%
P_3^3(t)$ completely fails at intermediate times and even goes through
singularity before returning to asymptotic value $\overline{x}$ at large
times.

In summary, in this case, the self-similar exponential approximant 
has been able to capture a highly non-trivial departure from a pure 
exponential relaxation which is dominating the relaxation process over 
a large time span, conditioned on the fact that the relaxation becomes 
asymptotically an exponential at long times. This shows again the power 
of this resummation method to capture significant deviations from 
exponentials in functions that belong to the exponential class in an 
asymptotic sense.

\section{Concluding Remarks}

In conclusion, we have demonstrated how the technique of self-similar
exponential approximants makes it possible to reconstruct exponential-type
functions, when only a few terms of their expansions are known.

This technique can also be applied to functions with asymptotic behavior
different from an exponential. In this case, by carefully
examining the convergence of the multipliers and of the approximants, 
it is possible to construct an accurate approximation for the sought 
function, while staying within the limits of applicability of the 
mean-field regime.

Further increase of accuracy will come from lifting the mean-field condition
used in deriving the self-similar exponential approximants. The result of
this approach will be presented elsewhere.

The Pad\'e approximants remain a valuable technique, but it has no much 
value in theories of relaxation and should be replaced by other 
techniques, possibly by the superexponentials presented here and their 
non mean-field extensions that we shall report in a future work.

\clearpage

Figure Captions

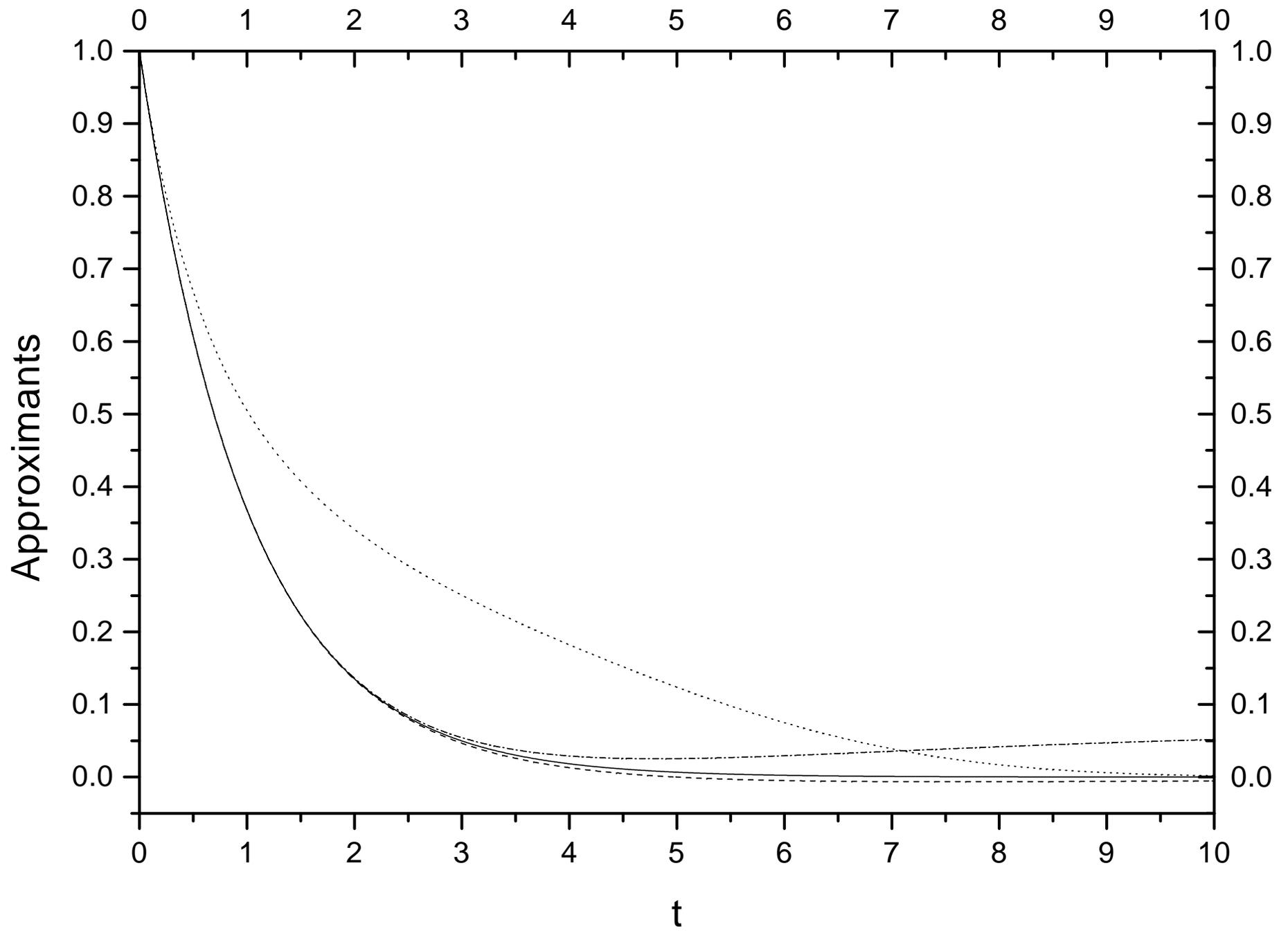
Figure 1. The conventional Pad\'e approximants, $P_3^2(t)$ (dash-dot) and $P_4^1(t)$
(dash) compared with $exp(-t)$ (solid). Approximant $\phi _{51}^{*}(t)$
(dot) is shown as well.

\vskip 0.5cm 
Figure 2. The relative percentage errors for the $P_3^2(t)$ (solid) and $P_4^1(t)\ $%
(dash) Pad\'e approximants, are shown. One should not be mislead by the
seemingly superior performance of $P_4^1(t)$, which is qualitatively wrong
in predicting negative values already for moderate times.
 
\vskip 0.5cm 
Figure 3. Illustration of the impact of ''noise''  $\eta $ (or $\theta )$ in the
coefficients of the power law expansion, onto the accuracy of the exponential
approximant $\phi _{22}^{*}$, characterized by the absolute error, $\phi
_{22}^{*}(t,\eta ,0)-\phi _{22}^{*}(t)\simeq b_1(t)\ \eta +b_2(t)\ \eta
^2+...$ and $\phi _{22}^{*}(t,0,\theta )-\phi _{22}^{*}(t)\simeq c_1(t)\
\theta +c_2(t)\ \theta ^2+$.  The absolute errors  in the lowest order,$%
\left| b_1(t)\right| $ (solid) and $c_1(t)$ (dash), are shown as functions
of time ( $\eta =\theta =1$).
 
\vskip 0.5cm 
Figure 4. Demonstration of the influence of the noise
introduced into the higher order terms in the expansion onto the absolute
error dependence on time.  The absolute errors in the lowest order,$\left|
b_1(t)\right| $ $\eta \ $(solid), the next order contribution to the error, $%
b_2(t)\ \eta ^2$ (dashed)$,$ and their sum (dotted line) are shown for $\eta
=0.1.$ Same notations as in Fig. 3.
 
\vskip 0.5cm 
Figure 5. It is impossible to distinguish the approximant $\phi _{55}^{*}$($t$)
(dash-dot) from the exact function, $\phi (t)=\sin (t)^2\exp (-t^2),$
intended to be shown with dots. Approximants $\phi _{41}^{*}(t)$ (solid
line),\ $\phi _{51}^{*}(t)$ (dash line) and Taylor expansions $\phi _{55}(t)$
(short dot) and $\phi _{44}(t)$ (dash-dot-dot)$\ $ are presented for
comparison as well.
 
\vskip 0.5cm 
Figure 6. The logarithms of the relative percentage errors for the Pad\'e
approximant $P_3^2(x)$ (dash) and for the super-exponential approximant $%
\phi _{55}^{*}(t)$ (solid), in the case of $\phi (t)=\sin (t)^2\exp (-t^2)$.
 
\vskip 0.5cm 
Figure 7. Multipliers $M_{22}^{*}(t)$ (dash-dot), $M_{33}^{*}(t)$ (dot), $%
M_{44}^{*}(t)$ (dash), $M_{55}^{*}(t)\ $(solid line)$\ $in the case of $\phi
(t)=\ln (1+t)\exp (-t^2)$ .
 
\vskip 0.5cm 
Figure 8. The self-similar exponential approximants $\phi _{44}^{*}(t)$ (dashed
line) and $\phi _{55}^{*}(t)$ (solid line), bracket the exact function $\phi
(t)=\ln (1+t)\exp (-t^2),$ shown with dash-dot-dot. The approximants $\phi
_{22}^{*}(t)$ (dash-dot line) and $\phi _{33}^{*}(t)$ (dotted line) are
presented as well.
 
\vskip 0.5cm 
Figure 9. The Pad\'e approximants P$_3^2(t)$ (dash), P$_4^1(t)$ (solid), P$_1^4(t)$
(dash-dot), P$_2^3(t)$ (dot) and $\phi _{55}^{*}(t)$ (short dot) are
compared to each other and with the exact $\phi (t)=\ln (1+t)\exp (-t^2)$
(dash-dot-dot)$.$
 
\vskip 0.5cm 
Figure 10. Logarithm of the relative percentage errors for the self-similar
exponential approximant $\phi _{55}^{*}(t)$ (solid line) and for the Pad\'e
approximant P$_1^4(t)$ (dash), in the case of $\phi (t)=\ln (1+t)\exp
(-t^2). $
 
\vskip 0.5cm 
Figure 11. The dependence of the multipliers $M_{22}^{*}(t)$ (dash-dot), $%
M_{33}^{*}(t)$ (dot), $M_{44}^{*}(t)$ (dash), $M_{55}^{*}(t)$ (solid )as a
function of time, in the case of $\phi (t)=\frac 1{1+t}\exp (-t)$
 
\vskip 0.5cm 
Figure 12. The approximants $\phi _{33}^{*}(t)$ (dot), $\phi _{44}^{*}(t)$ (dash), $%
\phi _{55}^{*}(t)$ (solid) are compared to the exact function $\phi
(t)=\frac 1{1+t}\exp (-t)$ (dash-dot).
 
\vskip 0.5cm 
Figure 13. The relative percentage errors for the Pad\'e approximant P$_3^2(x)$ (dash)
and for the self-similar exponential approximant $\phi _{55}^{*}(t)\ $%
(solid) are shown in the case of $\phi (t)=\frac 1{1+t}\exp (-t)$.
 
\vskip 0.5cm 
Figure 14. Dependence of the multipliers $M_{22}^{*}(t)$
(dash-dot), $M_{33}^{*}(t)$ (dot), $M_{44}^{*}(t)$ (dash), $M_{55}^{*}(t)\ $%
(solid line) as a function of time in the case of $\phi (t)=\exp (-t\ \left(
1+t\right) ^{-1/2}).$
 
\vskip 0.5cm 
Figure 15. The self-similar exponential approximants $\phi _{44}^{*}(t)$ (dashed
line) and $\phi _{55}^{*}(t)$ (solid line), bracket the exact function $\phi
(t)=\exp (-t\ \left( 1+t\right) ^{-1/2})$ shown with dash-dot-dot, rather
poorly. Approximants $\phi _{22}^{*}(t)$ (dash-dot line) and $\phi
_{33}^{*}(t)$ (dotted line) are presented as well.
 
\vskip 0.5cm 
Figure 16. Error of the approximant $\phi _{55}^{*}(t)$ (solid line) in the case
of $\phi (t)=\exp (-t\ \left( 1+t\right) ^{-1/2})$, compared with that of
the Pad\'e approximant $P_4^1(t),$ shown with dashed line.
 
\vskip 0.5cm 
Figure 17. Multipliers $M_{22}^{*}(t)$ (dash-dot), $M_{33}^{*}(t)$ (dot), $%
M_{44}^{*}(t)$ (dash), $M_{55}^{*}(t)\ $(solid line)$\ $in the case of the
Stieltjes function.
 
\vskip 0.5cm 
Figure 18. Approximants $\phi _{55}^{*}(t)$ (solid line) and $\phi _{44}^{*}(t)$
(dashed line) to the Stieltjes function, are compared with the exact
expression (dash-dot line) and with the Pad\'e approximant $P_3^2(t),$ shown
with dotted line.
 
\vskip 0.5cm 
Figure 19. In the case of Stieltjes function, it is shown here that
the Pad\'e approximant $P_3^2(t)\ $ shown with dashed line, easily
outperforms $\phi _{55}^{*}(t)$ shown with solid line
 
20. Illustration of how well the ``exact'' numerical solution to Eq.
(11), can be approximated by the superexponential approximants in
dimensionless units $X$ and $T$ , for $X(0)=50,\ m=0.85.\ $  The ``Exact''
numerical solution to the Eq. (11) is shown as the solid line; the second -order
approximation (15) is shown as the dash-dot-dot line; the third order approximation is
shown as the dash-dot line; the fourth order approximation is shown with dotted line,
while the best, fifth-order approximation, is shown with the dashed line and can
barely be distinguished from the exact solution. The naive exponential
expression $x_N(t)\ $(13) is shown with the short dash line.
 
\vskip 0.5cm 
Figure 21. Demonstration that the two-point Pad\'e approximant $P_3^3(t)$ (16)
shown with dashed line, completely fails at intermediate times. The approximate
solution (15), based on the fifth-order superexponential approximant and
shown with solid line, works well.

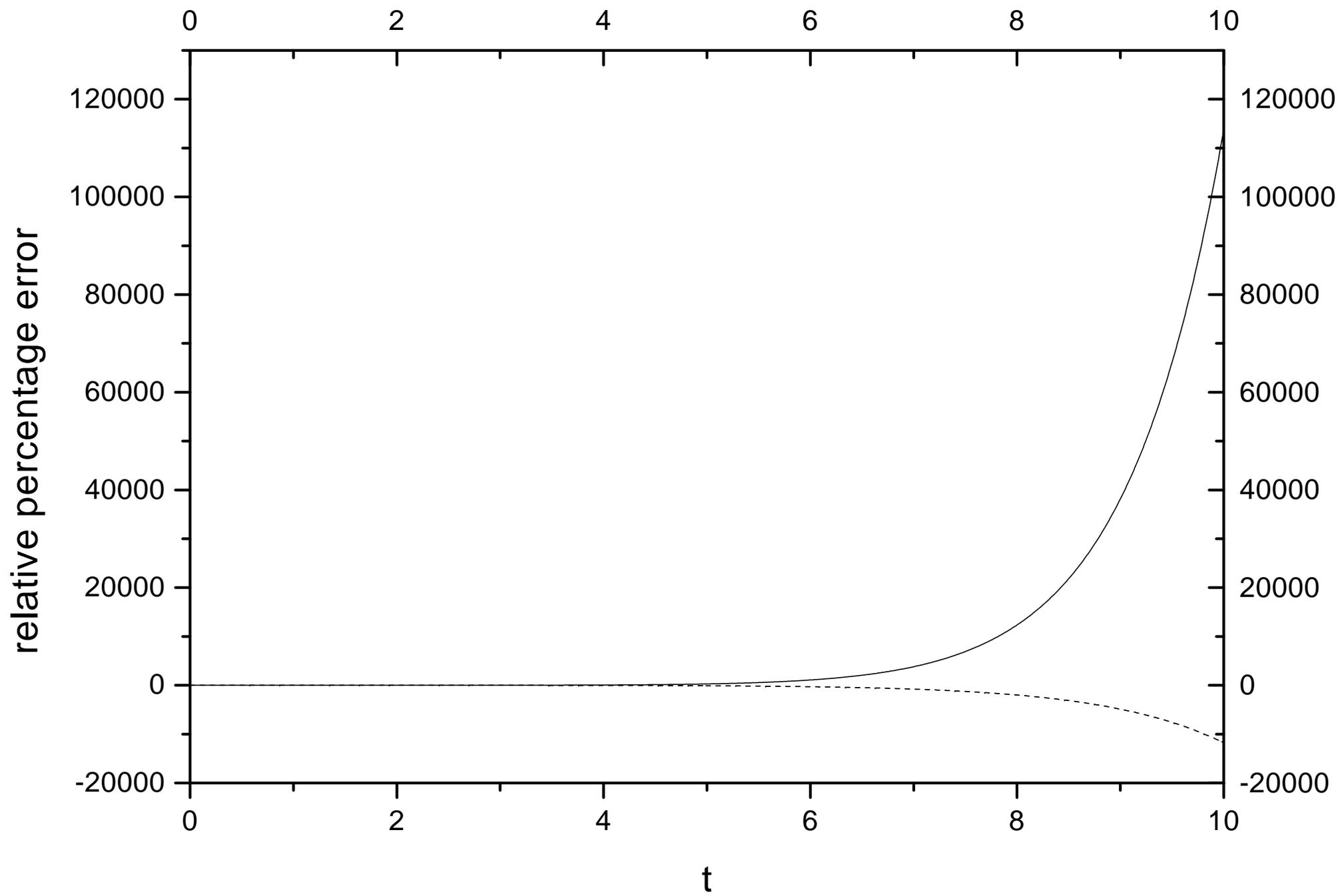
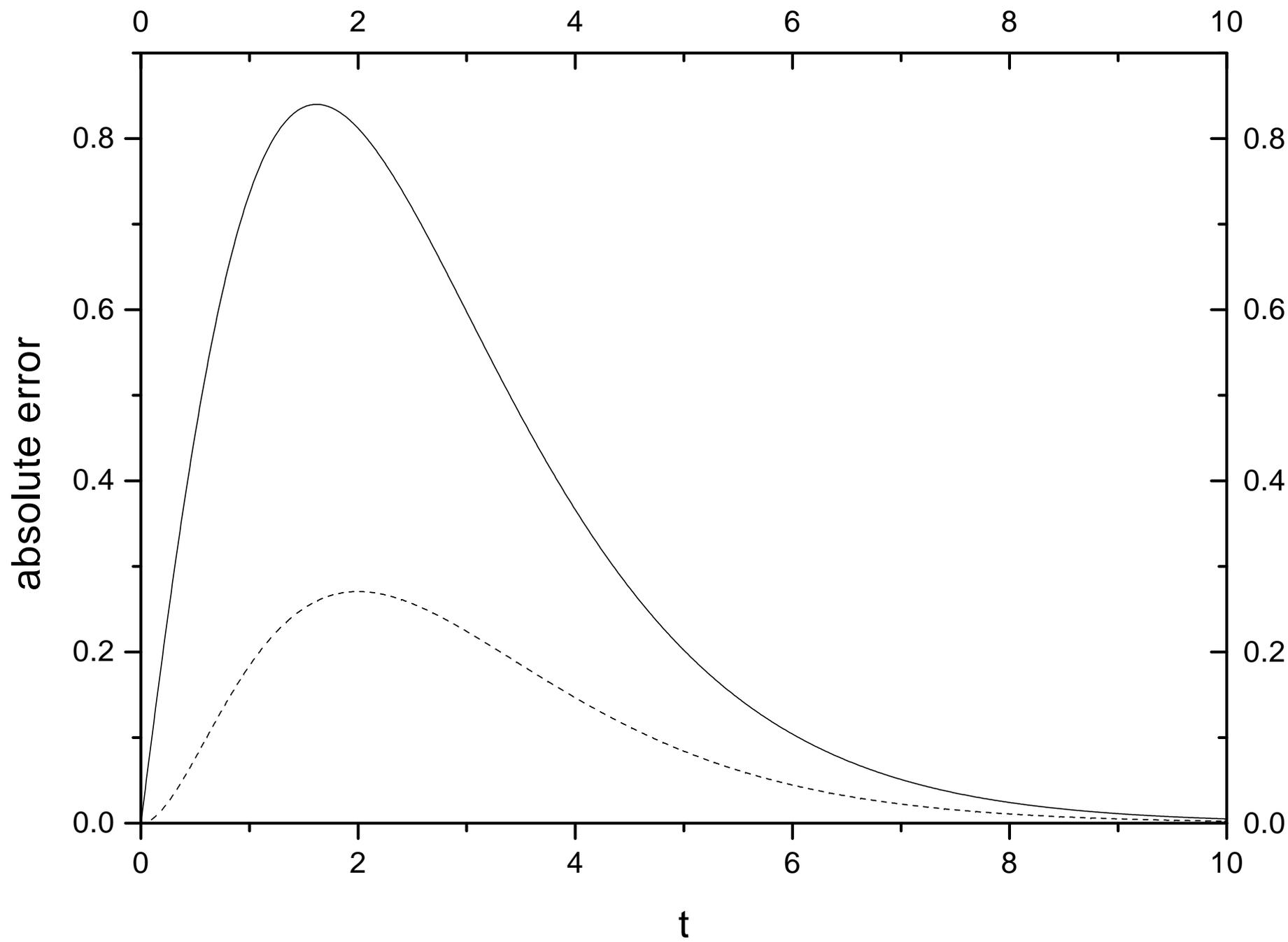
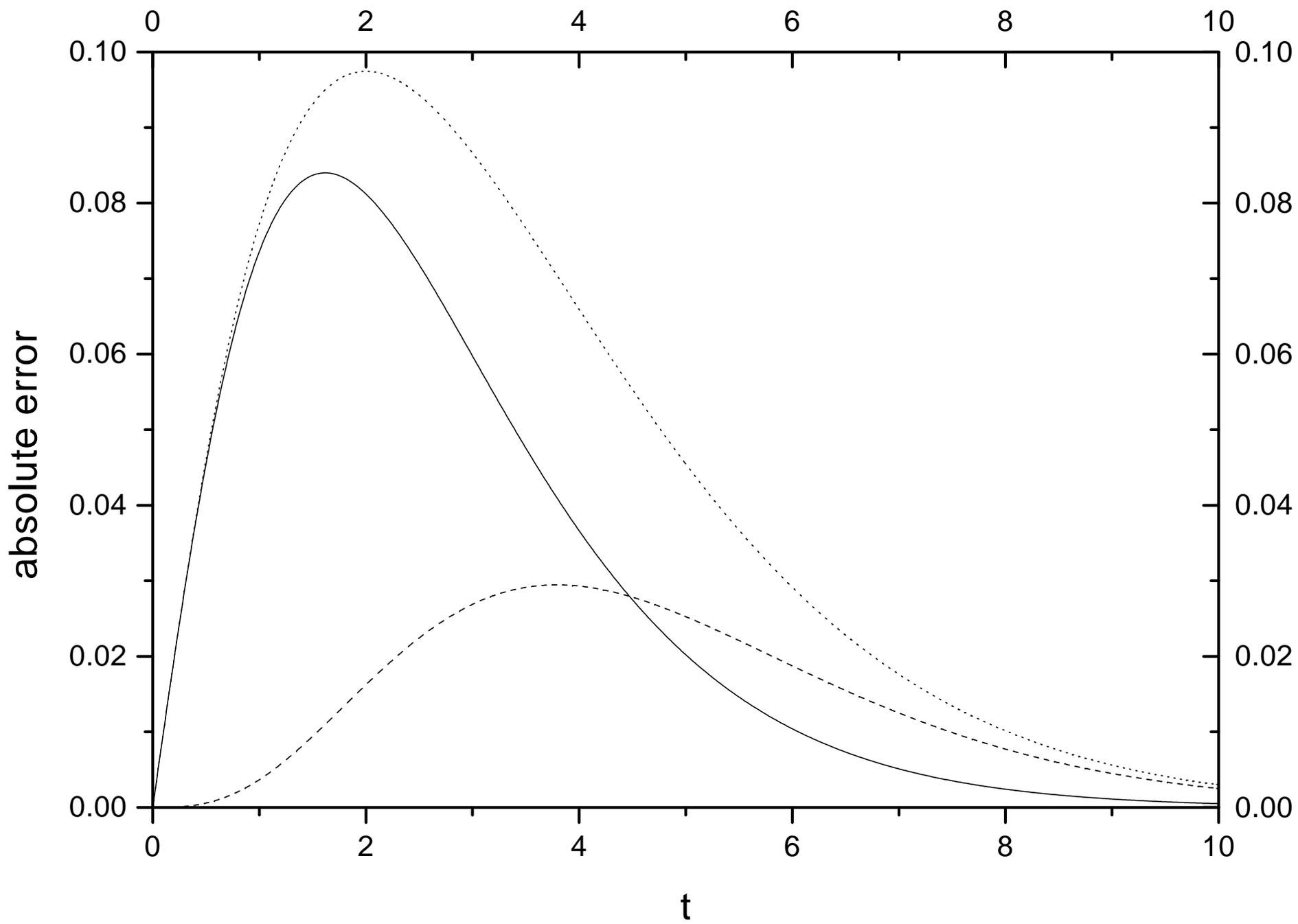
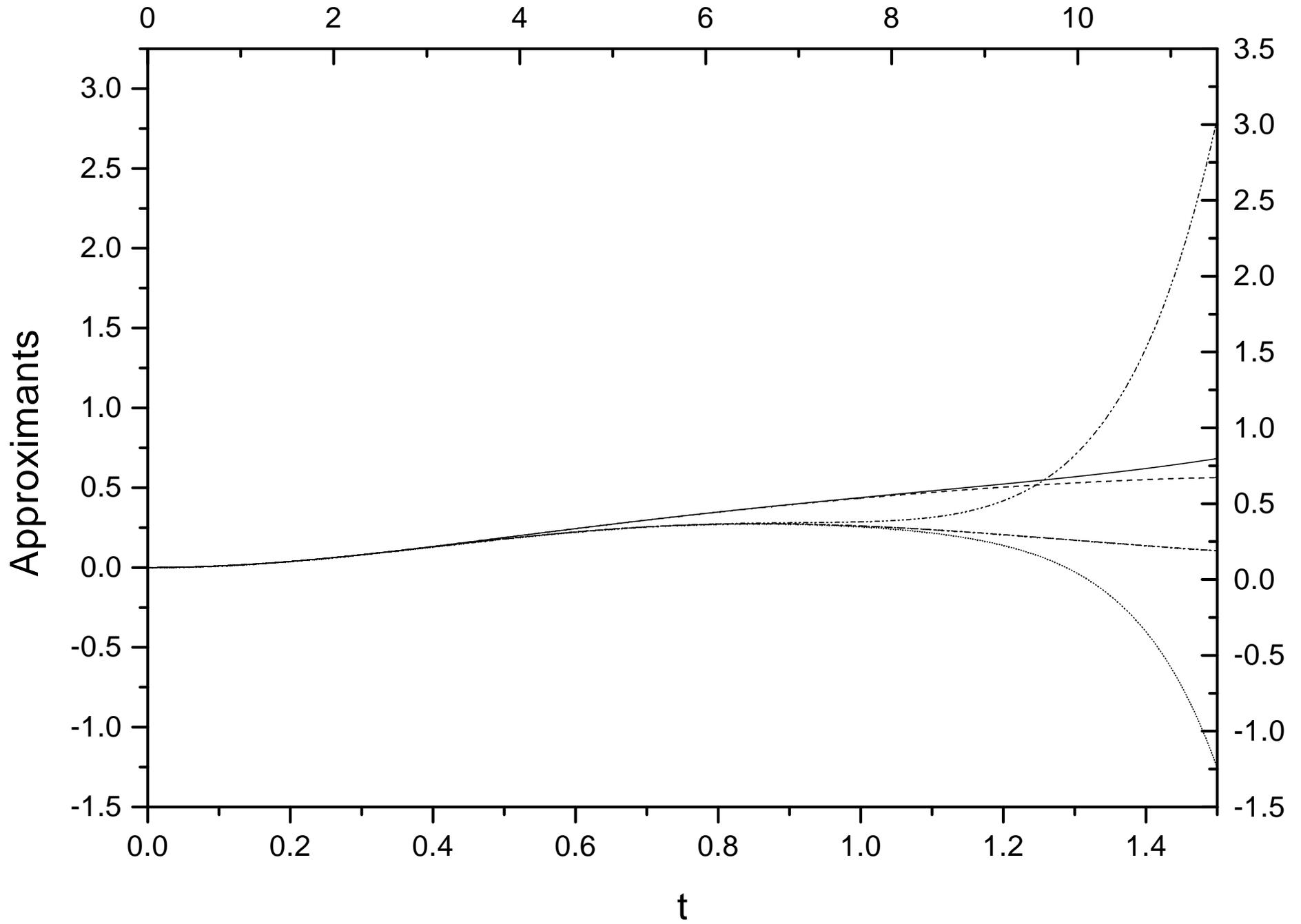
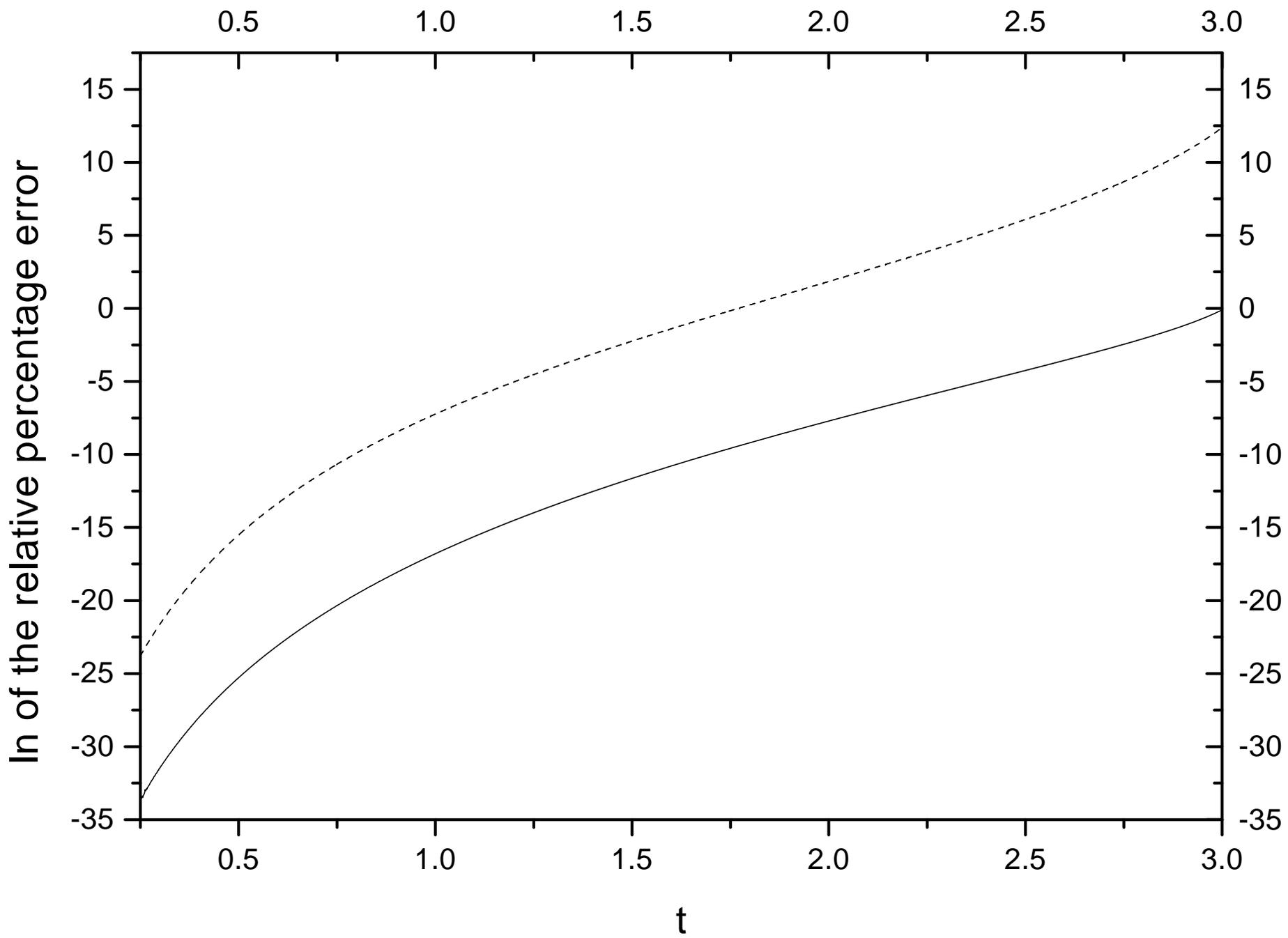
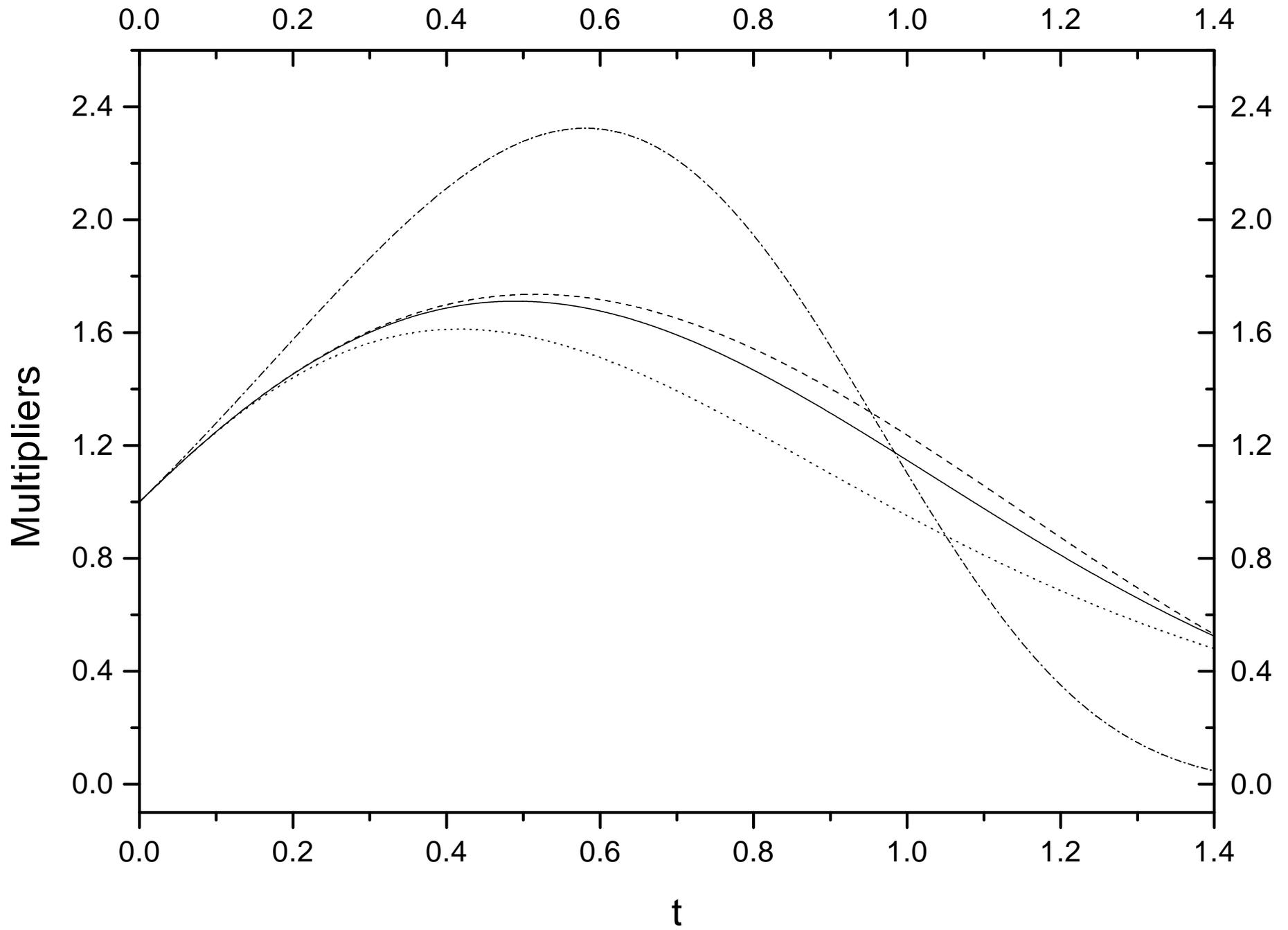
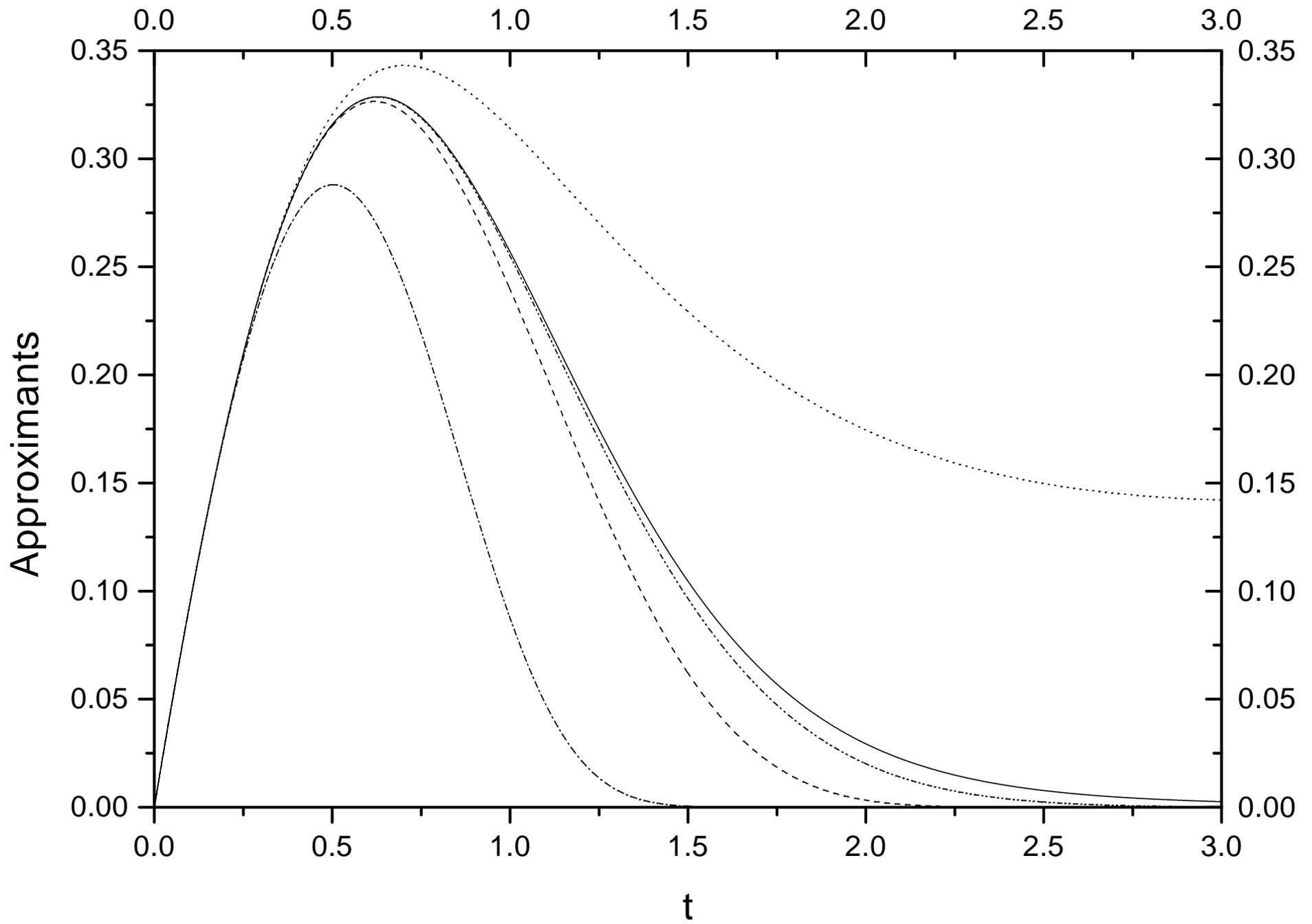
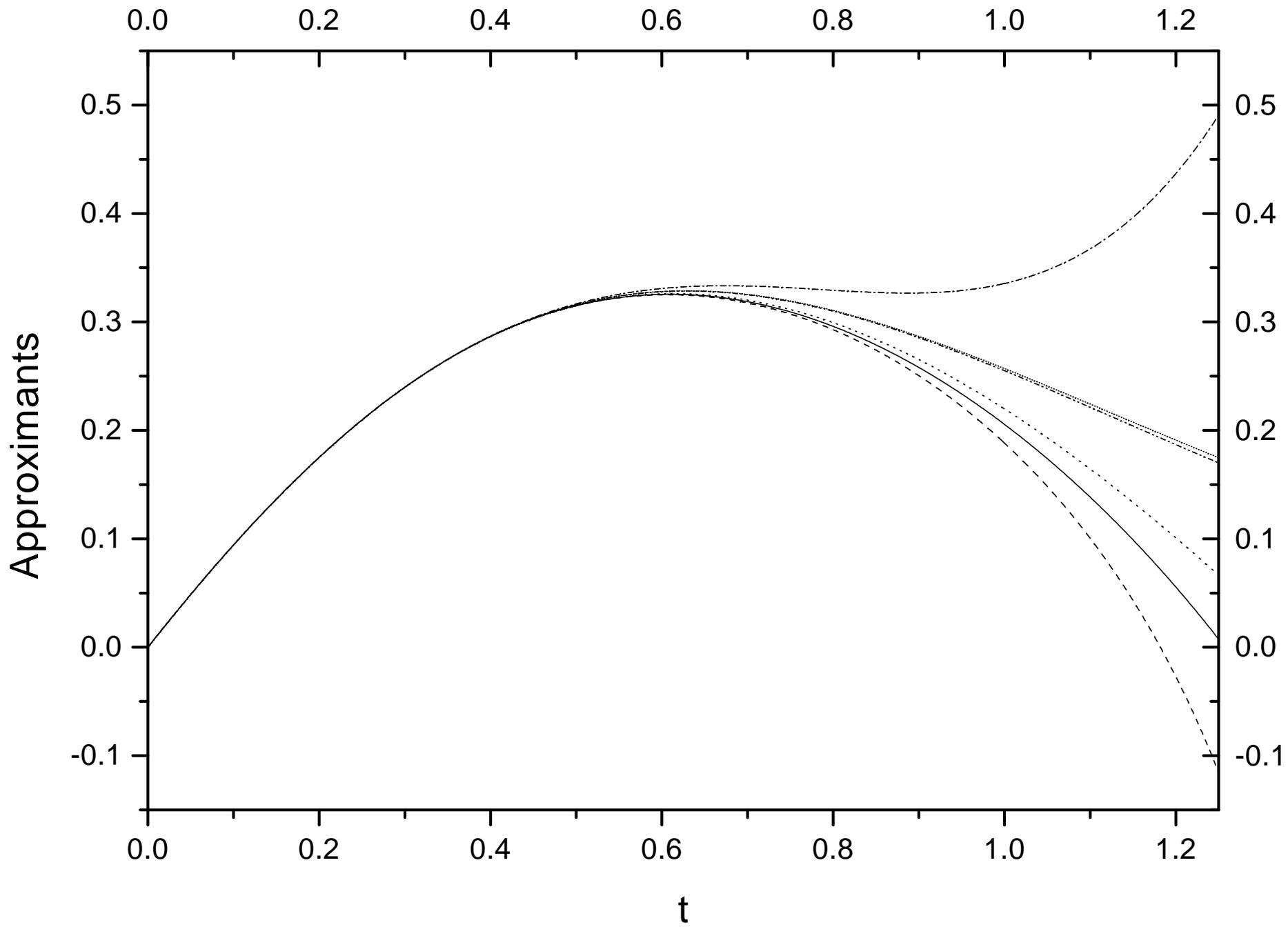
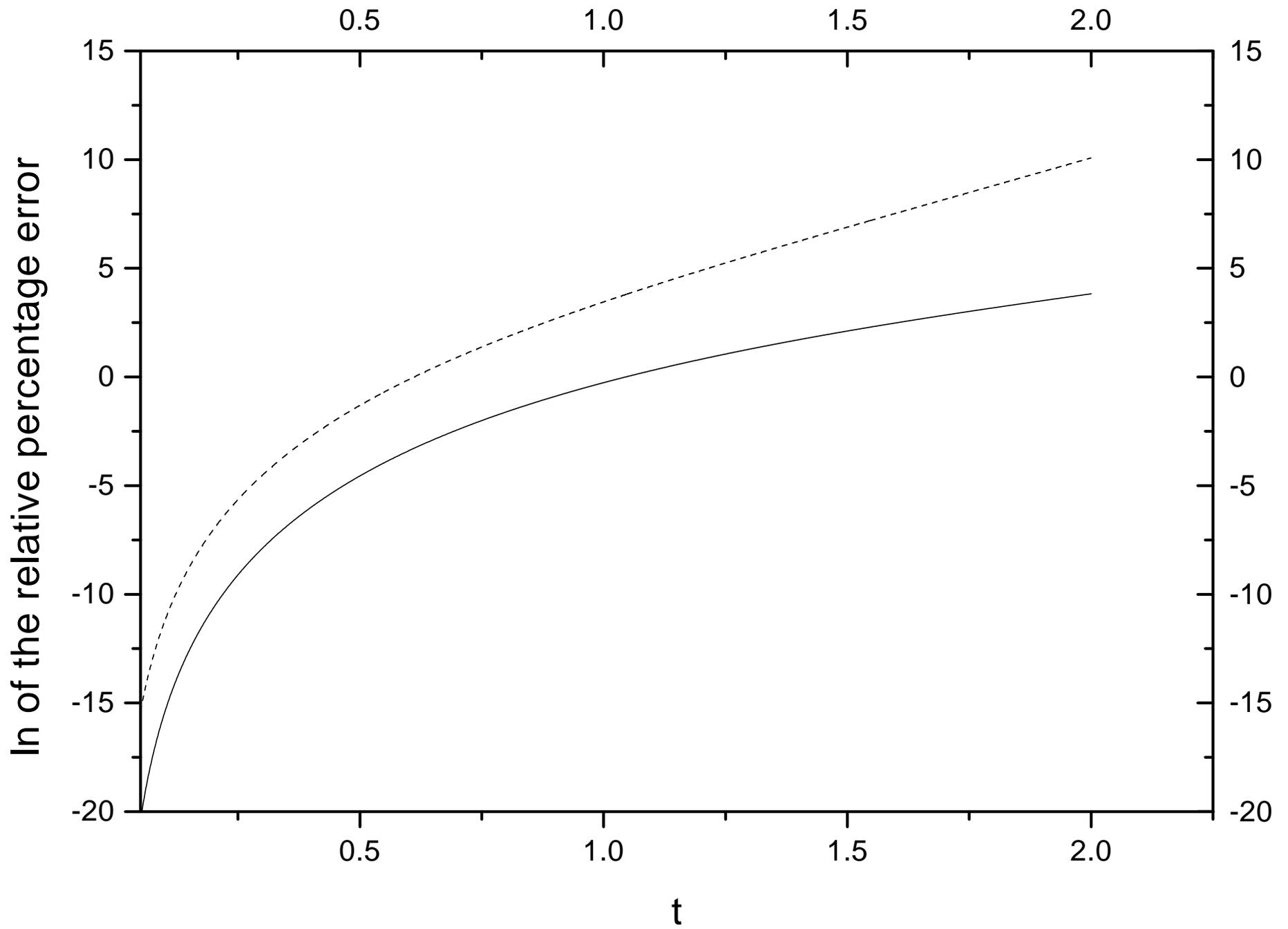
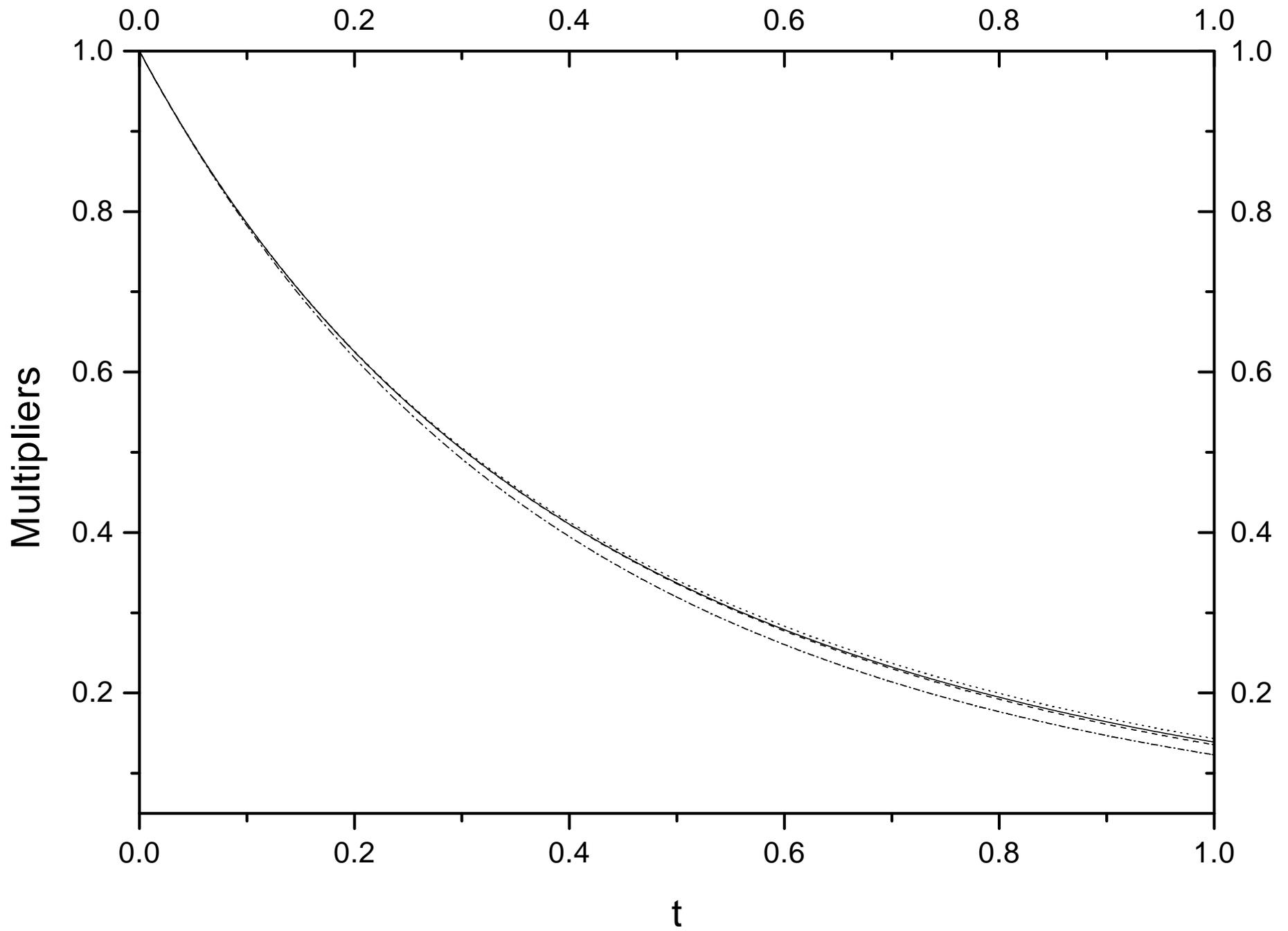
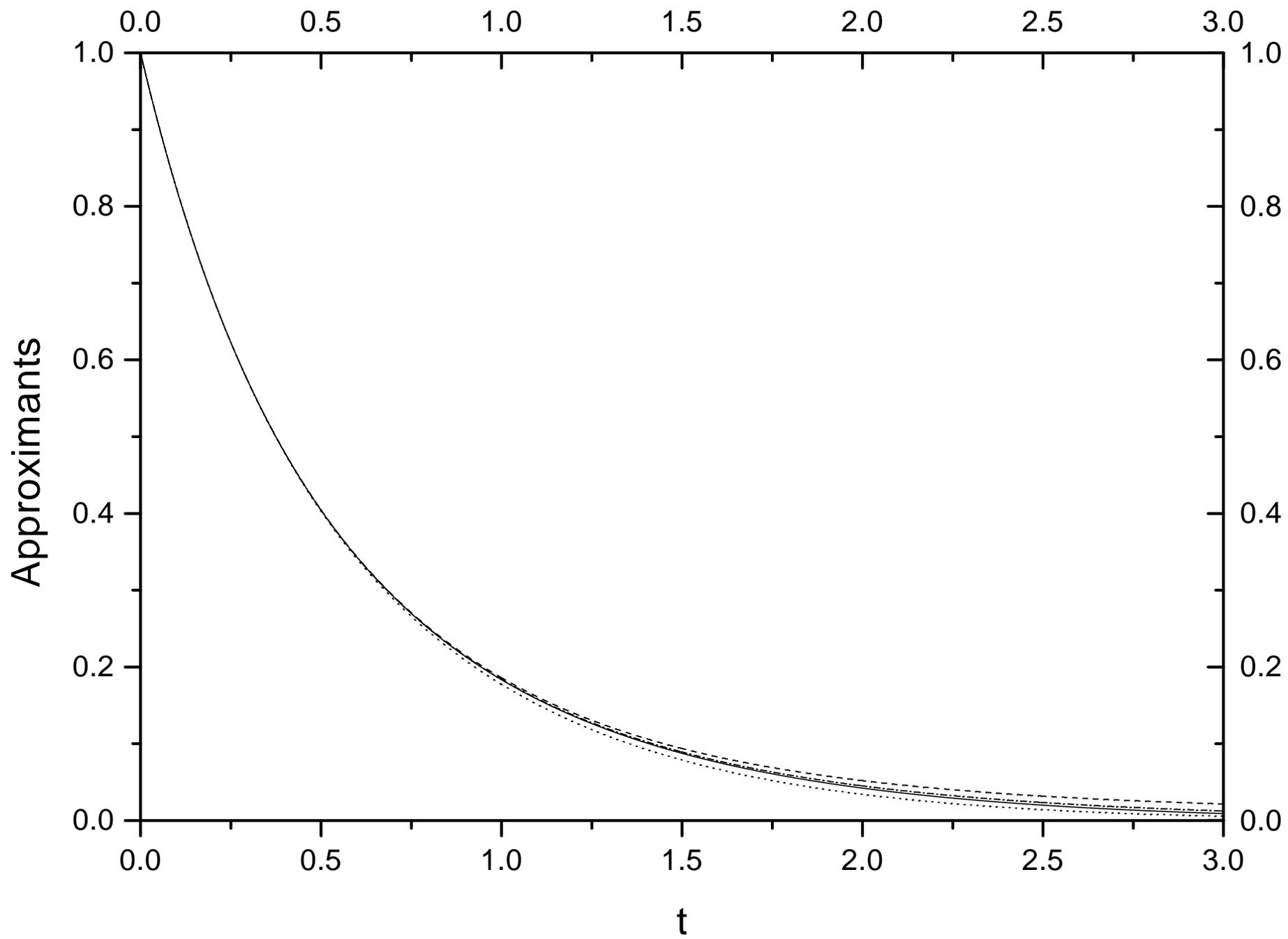
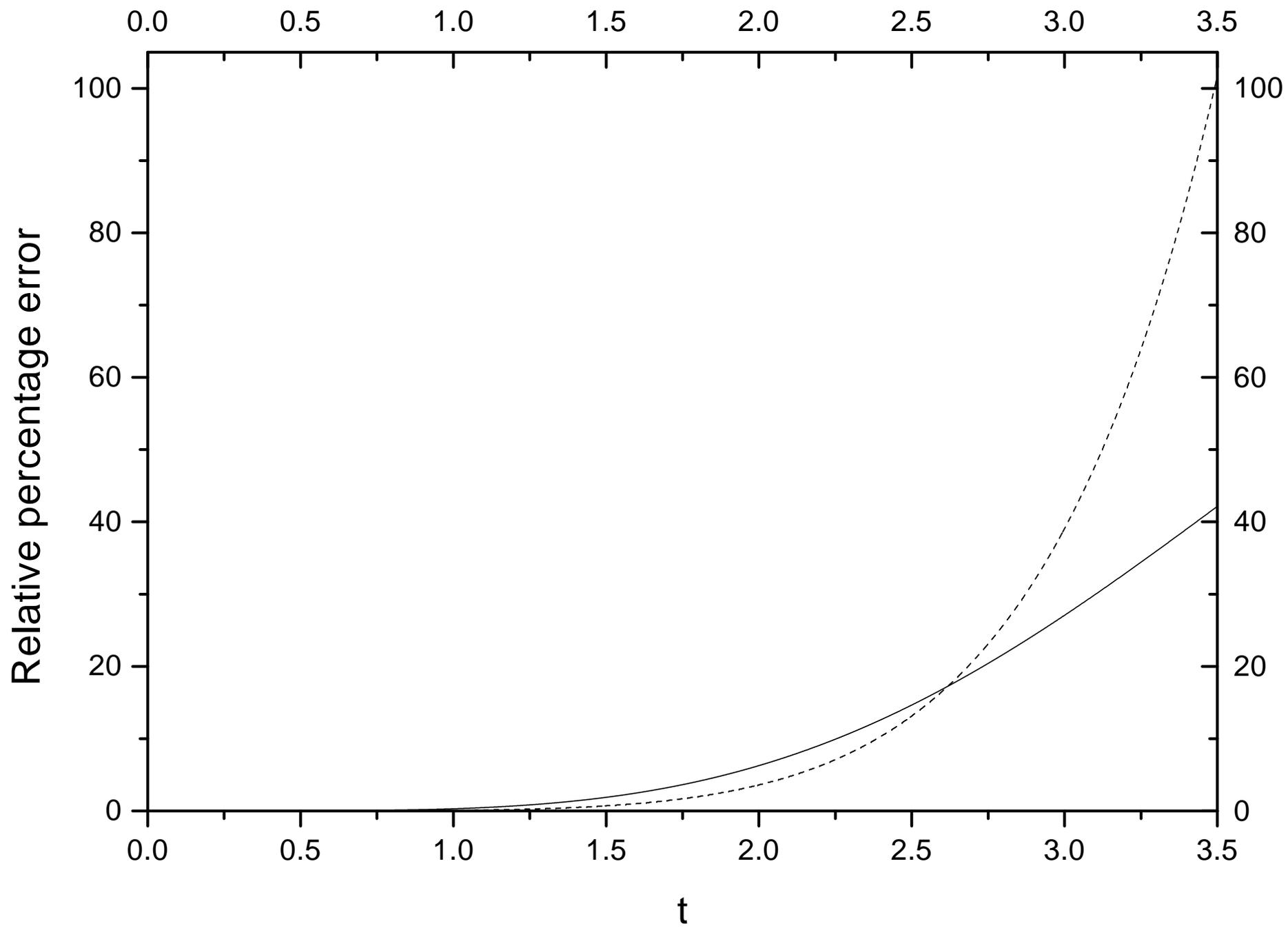
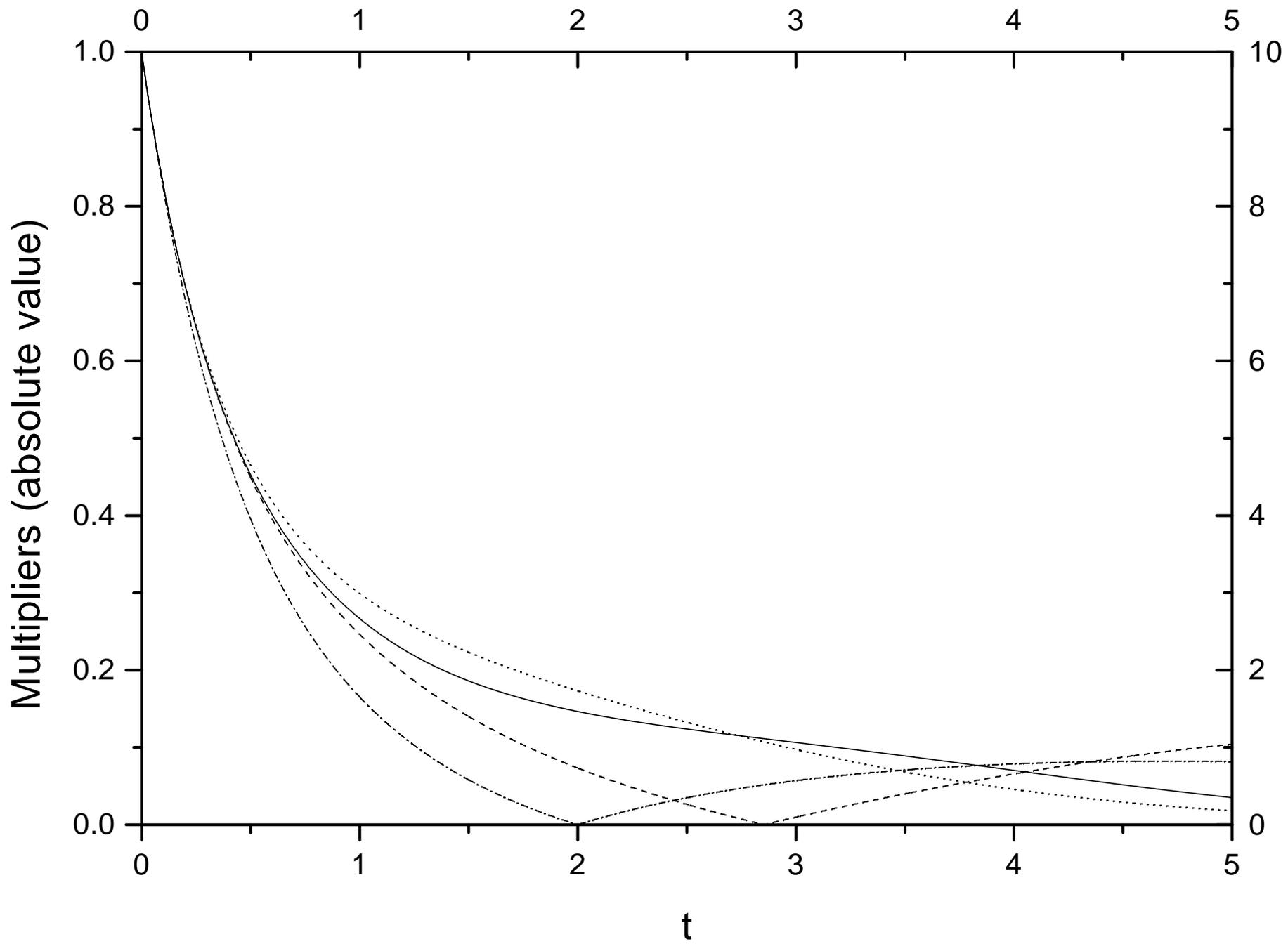
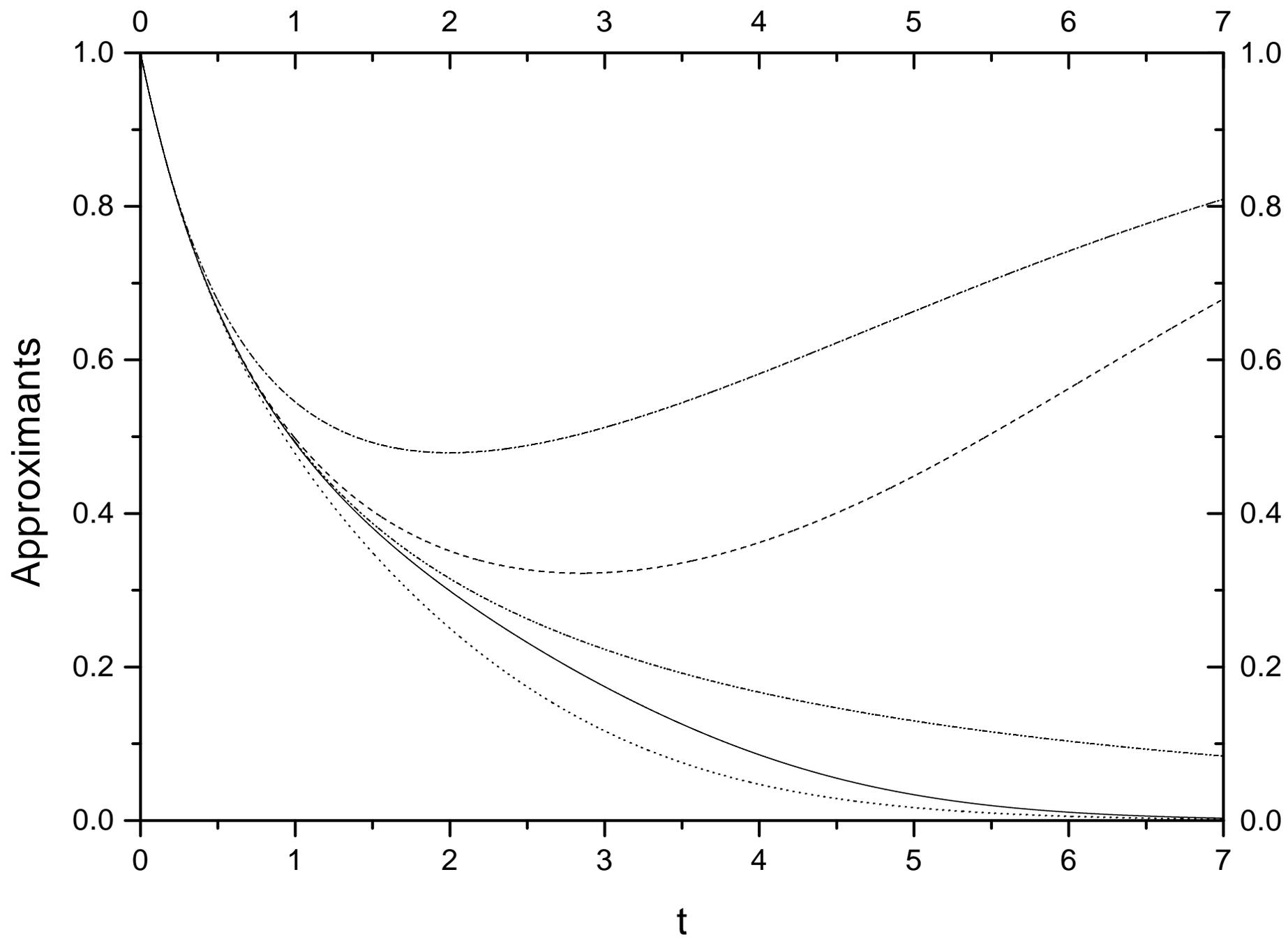
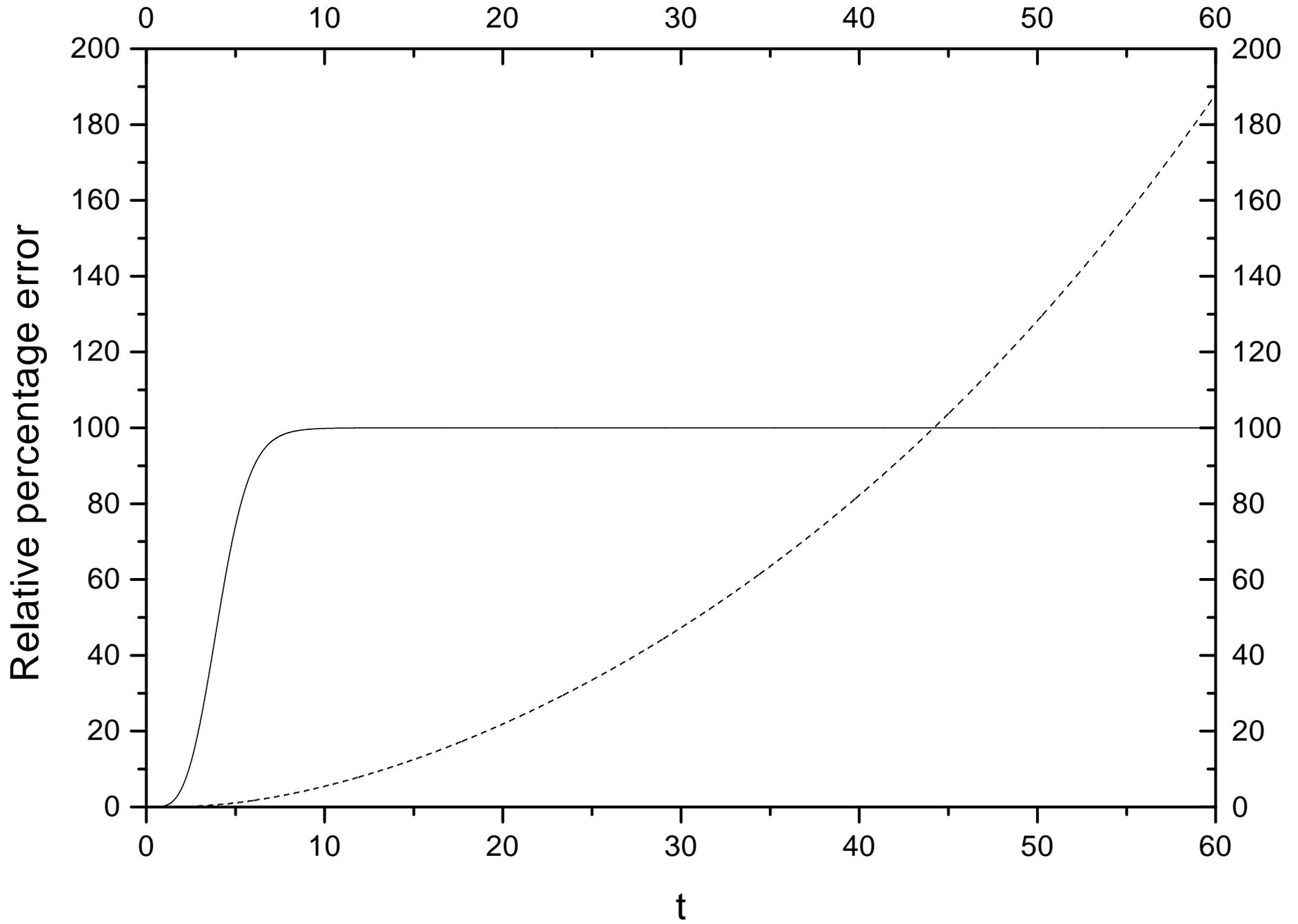
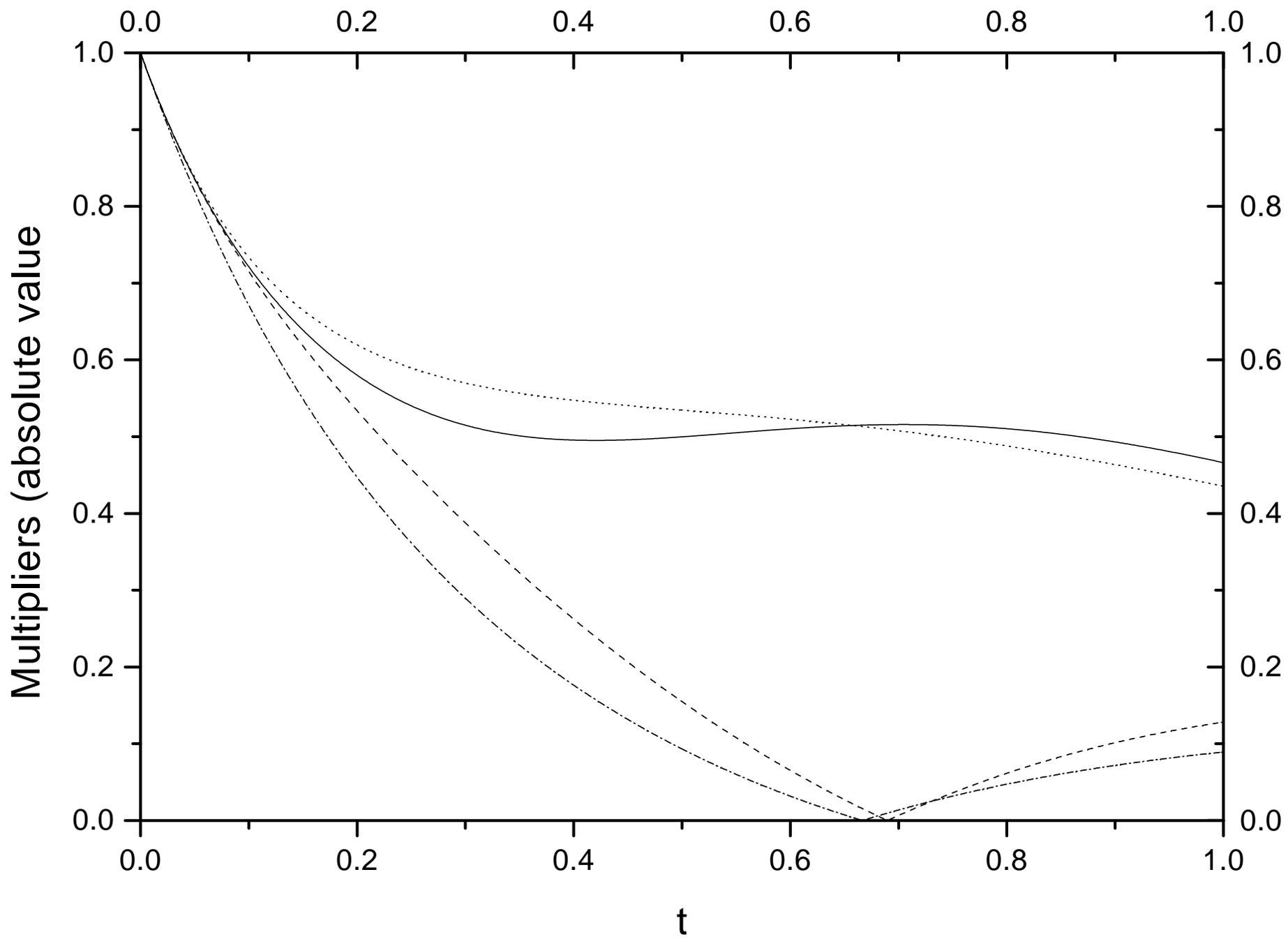
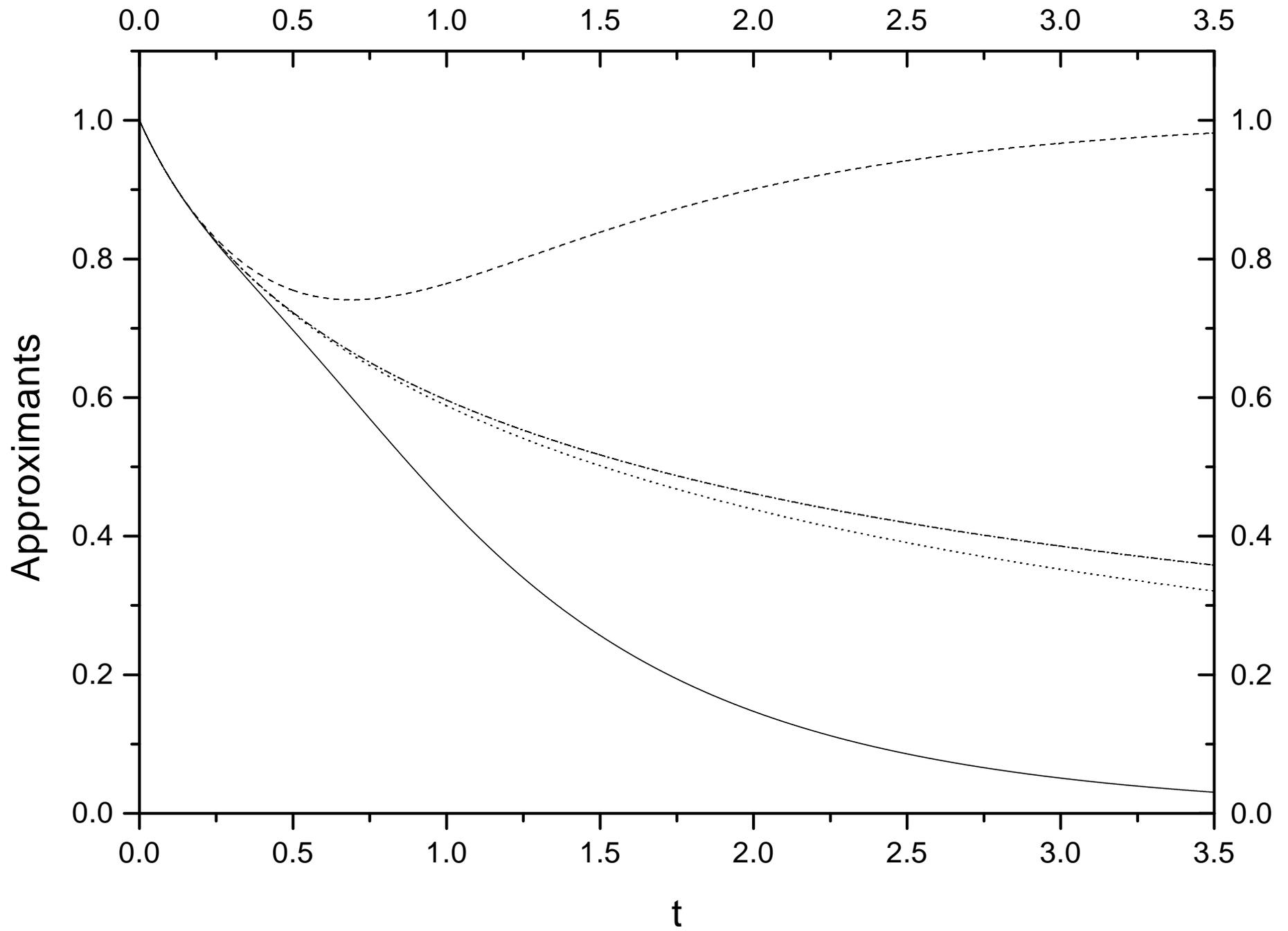
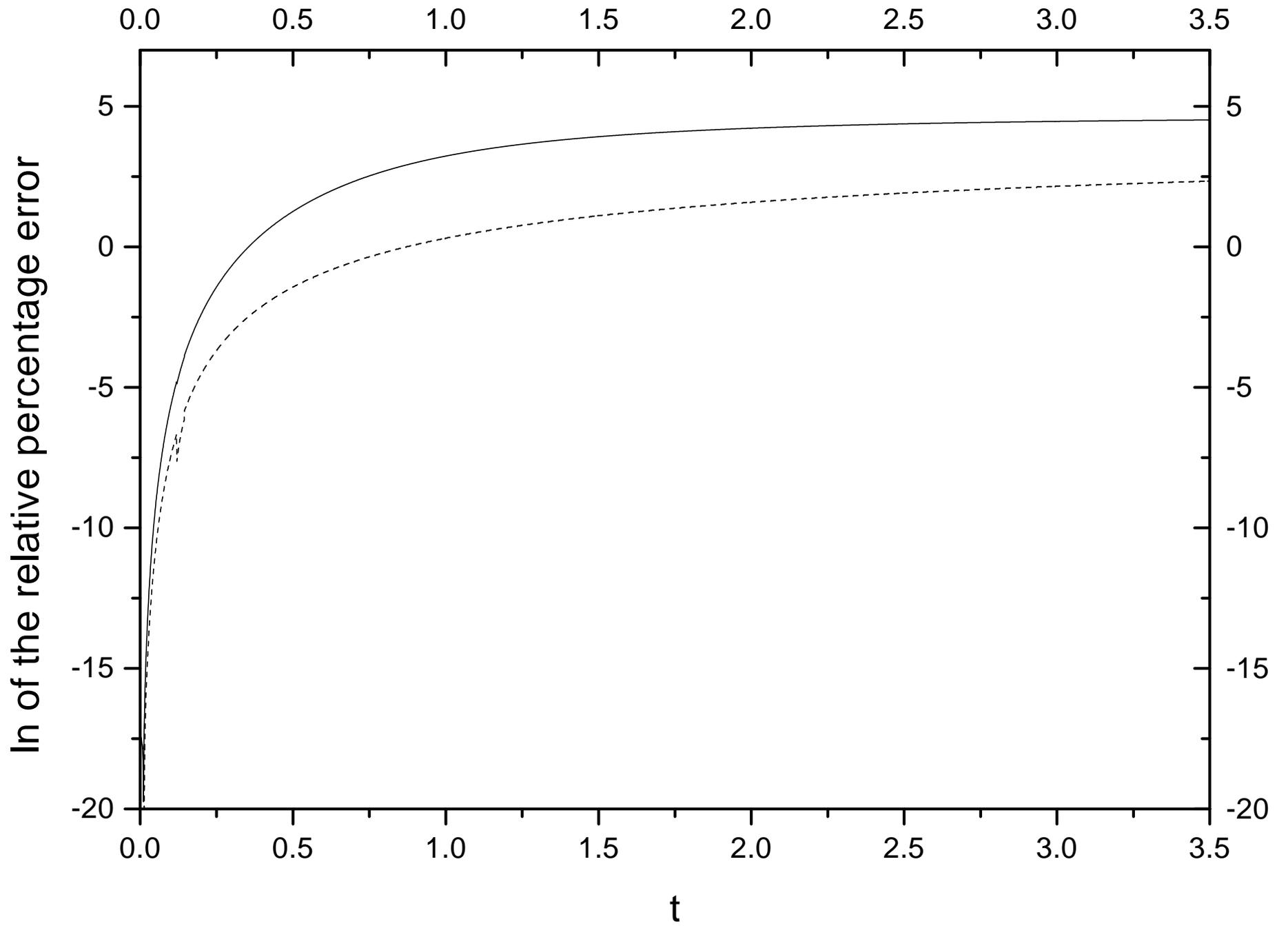
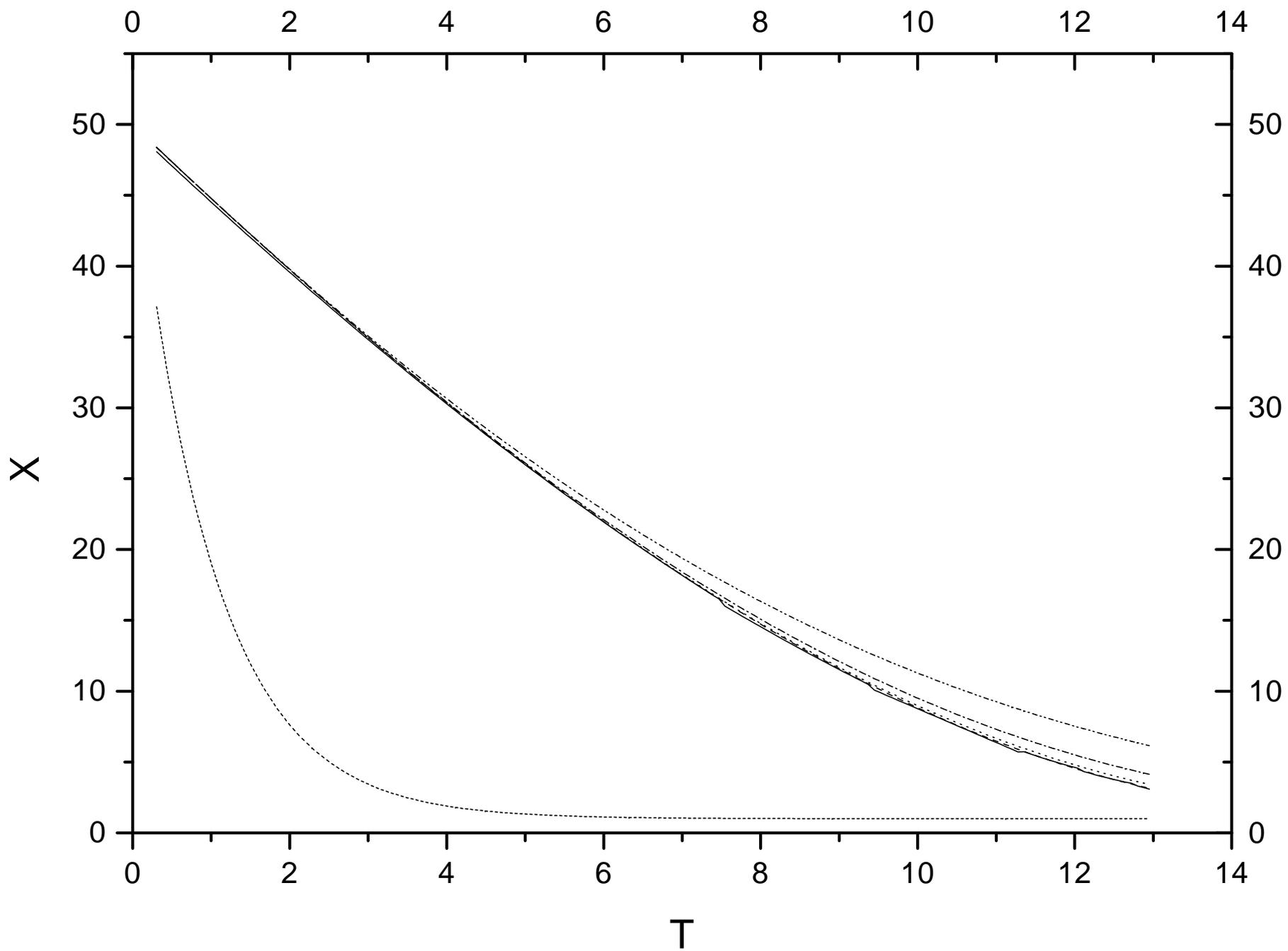
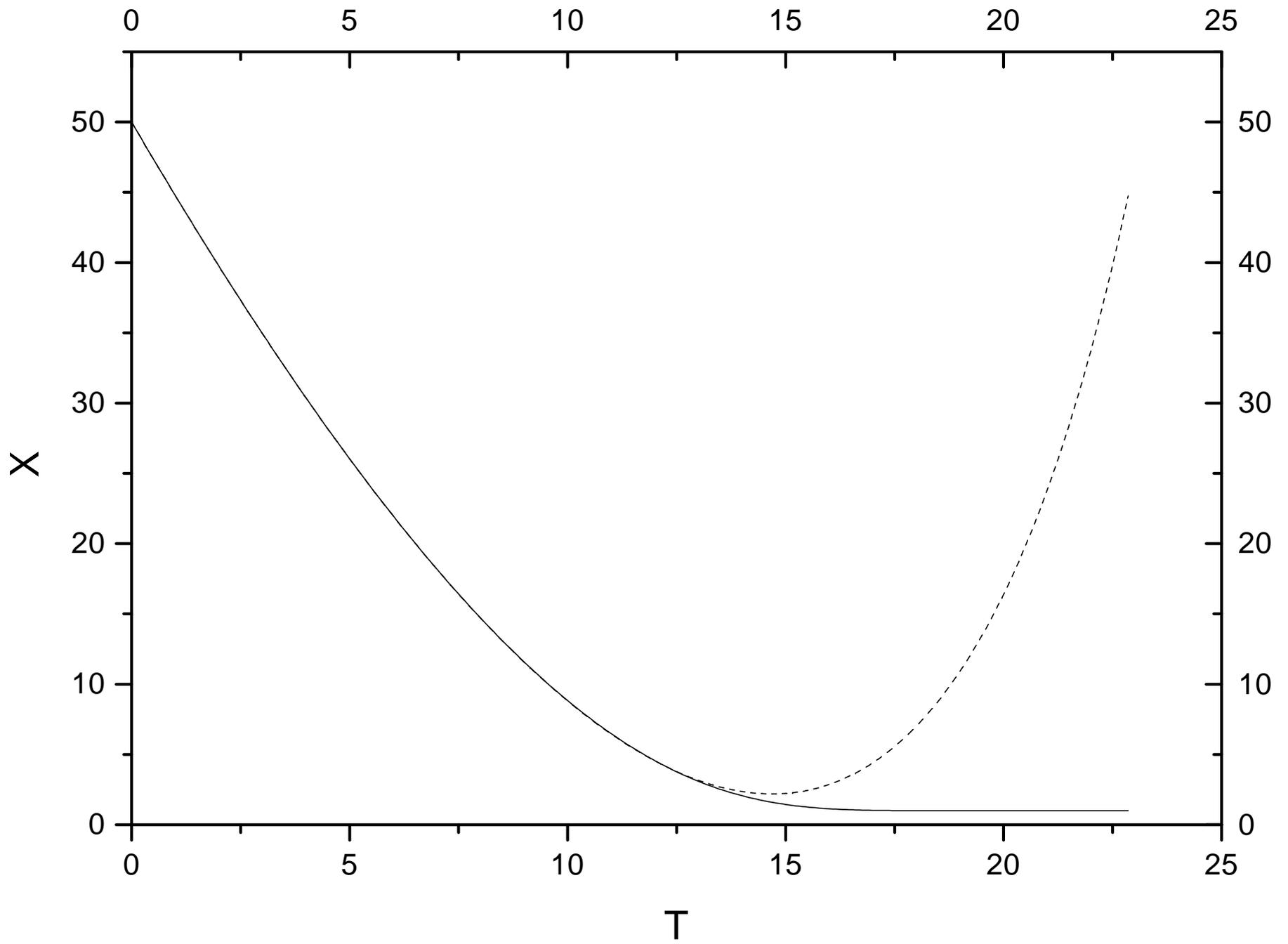
\end{document}